\documentclass[preprint,showpacs,preprintnumbers,amsmath,amssymb]{revtex4}
\pdfoutput=1

\usepackage{graphicx}% Include figure files
\usepackage{subfigure}
\usepackage{dcolumn}% Align table columns on decimal point
\usepackage{bm}% bold math
\usepackage{multirow, array} % para las tablas
\usepackage{float} % para usar [H]

\newcommand{\be}{\begin{equation}}
\newcommand{\en}{\end{equation}}
\newcommand{\bea}{\begin{eqnarray}}
\newcommand{\ena}{\end{eqnarray}}
\begin{document}

\title{Emergent Universe by Tunneling in a Jordan-Brans-Dicke Theory}
\author{Pedro Labra\~{n}a}
\email{plabrana@ubiobio.cl}
\author{Hobby Cossio}
\email{hcossio@ubiobio.cl}
 \affiliation{Departamento de F\'{i}sica, Universidad del
B\'{i}o-B\'{i}o and Grupo de Cosmolog\'{\i}a y Part\'{\i}culas Elementales UBB,
Casilla 5-C, Concepci\'on, Chile.}

\date{\today}% It is always \today, today,

\begin{abstract}

In this work we study an alternative scheme for an Emergent Universe
scenario in the context of a Jordan-Brans-Dicke theory, where the
universe is initially in a truly static state supported by a scalar
field located in a false vacuum. The model presents a classically
stable past eternal static state which is broken when, by quantum
tunneling, the scalar field decays into a state of true vacuum and
the universe begins to evolve following the extended open
inflationary scheme.

\end{abstract}

\maketitle

\section{Introduction}
\label{Int}

The standard cosmological model (SCM) \cite{weinberg,peebles,kolb}
and the inflationary paradigm \cite{Guth1,Albrecht,Linde1,Linde2}
are shown as a satisfactory description of our universe \cite{weinberg,peebles,kolb}. However, despite its great success,
there are still important open questions to be answered. One of
these questions is whether the universe had a definite origin,
characterized by an initial singularity or if, on the contrary, it
did not have a beginning, that is, it extends infinitely to the
past.

Theorems about spacetime singularities have been developed in the
context of inflationary universes, proving that the universe
necessarily has a beginning. In other words, according to these
theorems, the existence of an initial singularity can not be avoided
even if the inflationary period occurs, see Refs.~\cite{Borde:1993xh,Borde:1997pp, Borde:2001nh, Guth:1999rh,Vilenkin:2002ev}. In theses theorems it is demonstrated
that null and time-like geodesics are generally incomplete in
inflationary models, regardless of whether energy conditions are
maintained, provided that the average expansion condition ($H> 0$)
is maintained throughout of these geodesics directed towards the
past, where $H$ is the Hubble parameter.

The search for cosmological models without initial singularities has
led to the development of the so-called Emergent Universes models
(EU) \cite{Ellis:2002we,Ellis:2003qz, Mulryne:2005ef,
Mukherjee:2005zt,Mukherjee:2006ds,Banerjee:2007qi,Nunes:2005ra,Lidsey:2006md}.

In the EU scheme it is assumed that the universe emerged from a past
eternal Einstein Static (ES) state to the inflationary phase and
then evolves into a hot big bang era.
These models do not satisfy the geometrical assumptions of the
theorems \cite{Borde:1993xh,Borde:1997pp, Borde:2001nh, Guth:1999rh,Vilenkin:2002ev} and they provide specific examples
of non–singular inflationary universes.

\begin{figure}[t]
    \centering
    \includegraphics[scale=0.75]{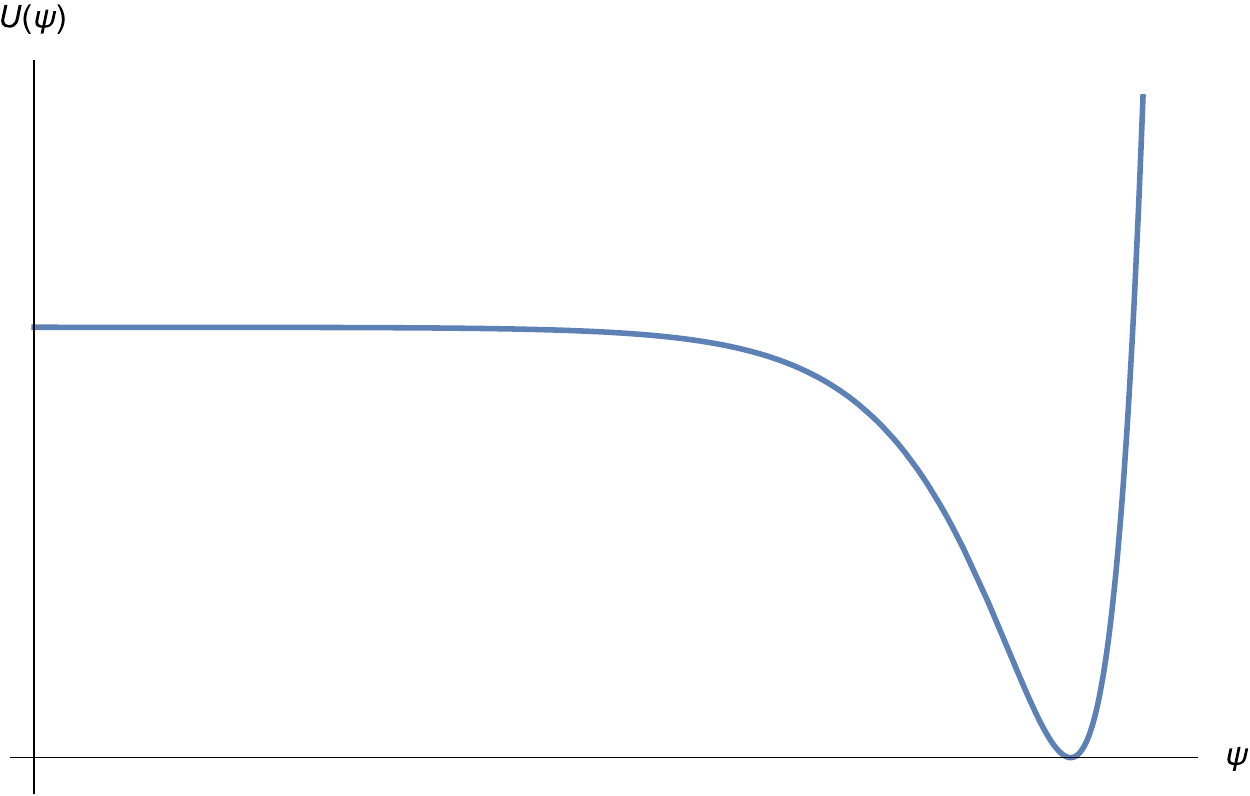}
    \caption{Schematic representation of a potential for a standard
Emergent Universe scenario.}
    \label{fig:Potential-1}
\end{figure}

Usually the EU models are developed by consider a universe dominated
by a scalar field, which, during the past-eternal static regime, is
rolling on the asymptotically flat part of the scalar potential (see
Fig.~(\ref{fig:Potential-1})) with a constant velocity, providing
the conditions for a static universe, see for example models
\cite{Ellis:2002we, Ellis:2003qz, Mulryne:2005ef,Nunes:2005ra,
Lidsey:2006md, delCampo:2007mp, delCampo:2009kp,
delCampo:2010kf,delCampo:2011mq,Guendelman:2011zza,Guendelman:2011fr,
Guendelman:2013dka, Guendelman:2014bva, Guendelman:2015uca,
delCampo:2015yfa}.
Other possibility is to consider EU models in which the scale factor
only asymptotically tends to a constant in the past
\cite{Mukherjee:2005zt, Mukherjee:2006ds,
Banerjee:2007sg,Debnath:2008nu, Paul:2008id,
Beesham:2009zw,Debnath:2011qi, Mukerji:2011wq, Labrana:2013oca, Huang:2015zma}.
We can note that in these schemes of Emergent Universe are not all
truly static during the static regime.

At this respect a new scheme for an EU model was proposed in Ref.~\cite{Labrana:2011np}, where the universe is initially in
a truly static state supported by a scalar field located in a false
vacuum, also see Refs.~\cite{Labrana:2013kqa,Labrana:2014yta}. The universe begins to evolve when, by quantum tunneling,
the scalar field decays into a state of true vacuum.
For simplicity, in this first approach to this new scheme of EU, the
model was developed in the context of General Relativity (GR).
In particular in Ref.~\cite{Labrana:2011np} was concluded that this new mechanism for an Emergent Universe is plausible and could be an interesting alternative  to the realization of the Emergent Universe scenario.

However, as this first model was developed in the context of
General Relativity, the past eternal static period, suffer from classical
instabilities associated with the instability of Einstein's static
universe.

The ES solution is unstable to homogeneous perturbations, as was early
discussed by Eddington in Ref.~\cite{Eddington} and more recently studied in
Refs.~\cite{Gibbons:1987jt, Gibbons:1988bm, Harrison:1967zz,
Barrow:2003ni}.
The instability of the ES solution ensures that any perturbations,
no matter how small, rapidly force the universe away from the static
state, thereby aborting the EU scenario.

This instability is possible to cure by going away from GR, for
example, by consider a Jordan-Brans-Dicke (JBD) theory at the
classical level, where it have been found that contrary to general
relativity, a static universe could be classically stable, see
Refs.~\cite{delCampo:2007mp, delCampo:2009kp, Huang:2014fia}.

\begin{figure}[t]
\centering
\includegraphics[scale=0.75]{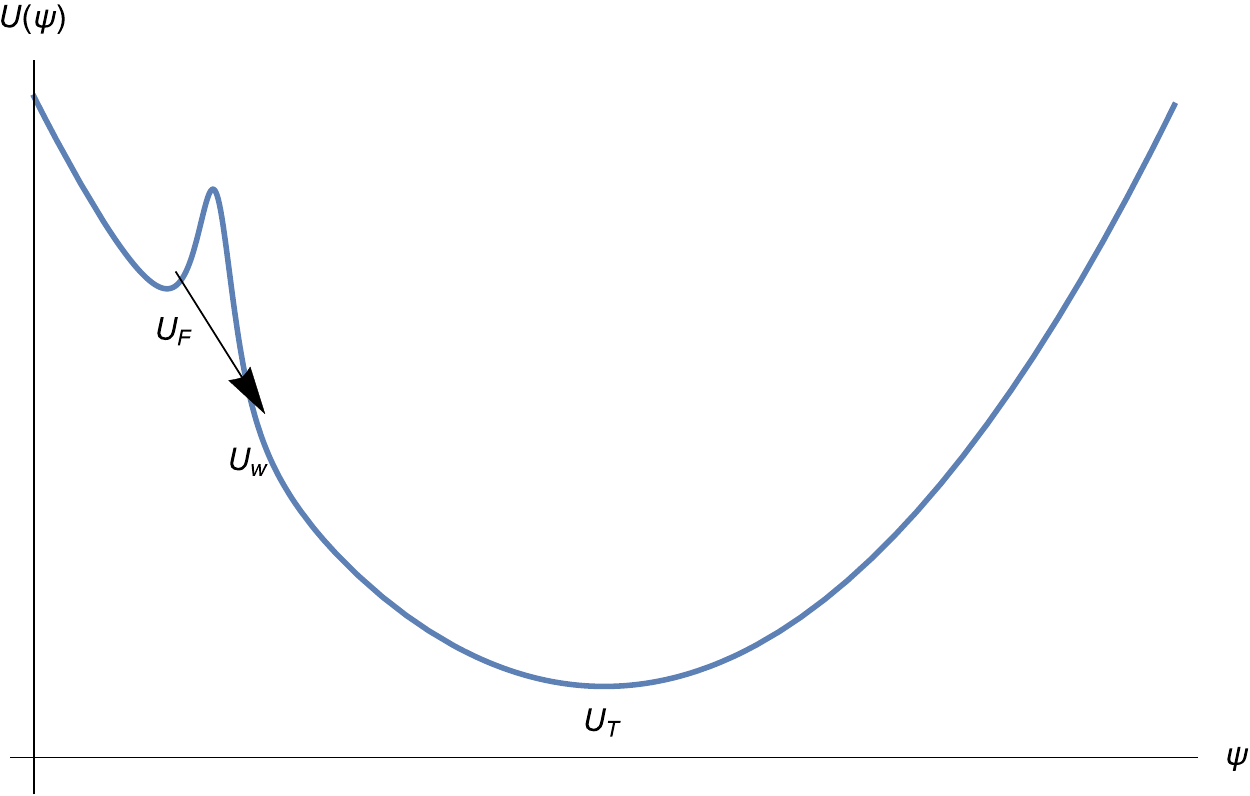}
\caption{Scalar field potential $U(\psi)$. Here
$U_F=U\left(\psi_F\right)$ and $U_T=U\left(\psi_T\right)$.}
\label{fig:Potential-2}
\end{figure}

In this work, we are interested in apply the scheme of Emergent
Universe by Tunneling of Ref.~\cite{Labrana:2011np} to EU
models which present stable past eternal static regimes.
In particular, we study this scheme in the context of a JBD theory,
similar to the one studied in Refs.~\cite{delCampo:2007mp,
delCampo:2009kp}, but where the static solution is supported by a
scalar field located in a false vacuum. In this case we are going to
show that, contrary to what happens in Ref.~\cite{Labrana:2011np}, the ES solution is classically stable.

The Jordan-Brans-Dicke \cite{Jbd} theory is a class of models in
which the effective gravitational coupling evolves with time. The
strength of this coupling is determined by a scalar field, the
so-called Brans-Dicke field, which tends to the value $G^{-1}$, the
inverse of the Newton's constant. The origin of Brans-Dicke theory
is found in the Mach's principle according to which the property of inertia of
material bodies arises from their interactions with the matter
distributed in the universe. In modern context, Brans-Dicke theory
appears naturally in supergravity models, Kaluza-Klein theories and
in all known effective string actions \cite{Freund:1982pg,
Appelquist:1987nr, Fradkin:1984pq, Fradkin:1985ys, Callan:1985ia,
Callan:1986jb, Green:1987sp}.

In particular in this work we are going to consider that the
universe is initially in a truly static state, which is supported by
a scalar field $\psi$ located in a false vacuum ($\psi = \psi_F$),
see Fig.~(\ref{fig:Potential-2}). The universe begins to evolve when,
by quantum tunneling, the scalar field decays into a state of true
vacuum. Then, a small bubble of a new phase of field value $\psi_W$
can be formed, and expands as it converts volume from high to low vacuum
energy and feeds the liberated energy into the kinetic energy of the
bubble wall. This process was first studied by Coleman
\& De Luccia in \cite{Coleman:1977py, Coleman:1980aw} in the context
of General Relativity.

If the potential has a suitable form, inflation and reheating may
occur inside the bubble as the field rolls from $\psi_W$
to the true minimum at $\psi_T$, in a similar way to what happens in
models of Open Inflationary Universes and Extended Open Inflationary
Universes, see for example \cite{linde, re8, delC1, delC2,
Balart:2007je}, where the interior of the bubble
is modeled by an open Friedmann-Robertson-Walker universe.

%%%%%%%%%%%%%%%%%%%%%%%%%%

\begin{figure}[t]
    \begin{center}
        \includegraphics[scale=0.65]{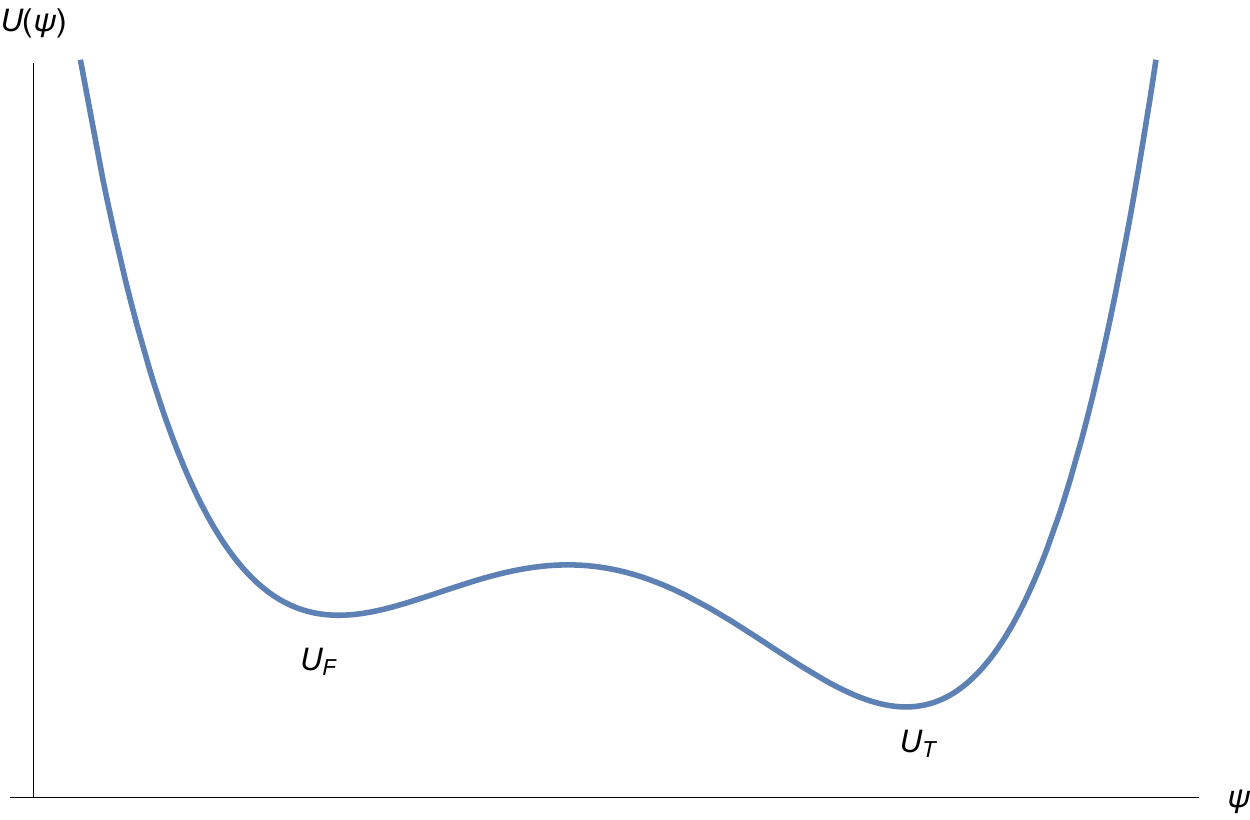}
        \caption{Potential with a false and true vacuum.}
        \label{fig:Potential-3}
    \end{center}
\end{figure}

%%%%%%%%%%%%%%%%%%%%%%%%%%%%%%%

The advantage of the EU by Tunneling scheme (and of the Emergent
Universe in general), over the Eternal Inflation scheme is that it
correspond to a realization of a singularity-free inflationary
universe.
As was discussed in Refs.~\cite{Borde:1993xh,Borde:1997pp,
Borde:2001nh, Guth:1999rh,Vilenkin:2002ev}, Eternal Inflation is
usually future eternal but it is not past eternal, because in
general space-time that allows for inflation to be future eternal,
cannot be past null complete. On the other hand Emergent Universe
are geodesically complete.

%%%%%%%%%%%

Notice that in the EU by Tunneling scheme, the metastable state
which support the initial static universe could exist only a finite
amount of time. Then, in this scheme of Emergent Universe, the
principal point is not that the universe could have existed an
infinite period of time, but that in theses models the universe is
non-singular because the background where the bubble materializes is
geodesically complete.

This implies that we have to consider the problem of the initial
conditions for a static universe. Respect to this point, there are
very interesting possibilities discussed for example in the early
works on EU \cite{Ellis:2003qz} and more recently in \cite{Labrana:2011np}.
One of these options is to explore the possibility of an Emergent
Universe scenario within a string cosmology context \cite{Antoniadis}. Other possibility is that the initial
Einstein Static universe is created from "nothing" \cite{Tryon,Vilenkin-cre}, see Refs.~\cite{Mithani} for explicit examples.
It is interesting to mention that the study of the Einstein Static solution as a preferred initial state for our universe have been considered in the past,
where it has been proposed that entropy considerations favor the ES
state as the initial state for our universe \cite{Gibbons:1987jt,Gibbons:1988bm}.

In this paper we consider a simplified version of this scheme, with the focus
on studying the process of creation and evolution of a bubble of
true vacuum in the background of an ES universe in the context of a
JBD theory.
This is motivated because we are mainly interested in the study of
new ways of leaving the static period and begin the inflationary
regime for Emergent Universe models which present a classically
stable static state period.

In particular, in this paper we consider a JBD theory where one of the matter content
of the model is a scalar field (inflaton) with a potential
similar to Fig.~(\ref{fig:Potential-3}) and we study the process of
tunneling of the scalar field from the false vacuum $U_F$ to the
true vacuum $U_T$ and the subsequent creation and evolution of a
bubble of true vacuum in the background of an stable ES universe.
The simplified model studied here contains the essential elements of
the scheme we want to present, so we postpone the detailed study of
the inflationary period, which occurs after the tunneling, for
future work.

The paper is organized as follow. In Sect.~\ref{sec:estadoestatico}
we study a Einstein static universe supported by a scalar field
located in a false vacuum and its stability in the context of a JBD
theory. In Sect. \ref{sec:tunel} we study the tunneling process of
the scalar field from the false vacuum to the true vacuum and the
subsequent creation of a bubble of true vacuum in the background of
the Einstein static universe for a JBD theory. In
Sect.~\ref{sec:evolucionbur} we study the evolution of the bubble
after its materialization. In Sect.~\ref{sec:conclu} we summarize
our results.

\section{False Vacuum and the ES State in JBD Theories}\label{sec:estadoestatico}

In this paper we consider a scheme for EU scenario where the universe is initially in a classically stable static state. This state is supported by a scalar field located in a false vacuum. The universe begins to evolve when, by quantum tunneling, the scalar field decays to a state of true vacuum. For this reason we will begin by studying the possibility of obtain a static and classically stable ES solution in this theory when the scalar field is in a false vacuum.

We consider the following JBD action for a self-interacting potential and matter, given by \cite{Green:2012oqa}
\begin{equation} \label{accion}
S=\int d^4x \sqrt{-g}\left[\frac{1}{2}\phi{\cal R}- \frac{1}{2}\frac{\omega}{\phi}\left(\nabla \phi\right)^2 + V(\phi)+
\left(\nabla \psi\right)^2 - U(\psi) + {\cal L}_m \right],
\end{equation}
where ${\cal R}$ is the Ricci scalar curvature, $\phi$ is the JBD field, $\omega$ is the JBD parameter, $V(\phi)$ is the potential associated to the JBD field, $\psi$ is the scalar  field (inflaton), $U(\psi)$ is the scalar potential and ${\cal L}_m$ denotes the Lagrangian density of a barotropic perfect fluid. In this theory $1/\phi$ plays the role of the gravitational constant, which changes with time. This action also matches the low energy string action for $\omega=-1$, see \cite{Green:1987sp}.

Following the EU scheme we consider a closed Friedmann-Robertson-Walker metric:
\begin{equation}\label{metrica}
    ds^2 = dt^2-a(t)^2\left[\frac{dt^2}{1-r^2} + r^2 \left(d\theta^2 + \sin^2 \theta d\varphi^2 \right)\right],
\end{equation}
where $a(t)$ is the scale factor and $t$ represents the cosmological time. The content of matter is modeled by a standard perfect fluid with an effective state equation given by $P_f=\left(\gamma-1\right)\rho_f$, with $\gamma$ constant, and a scalar field (inflaton) for which
\begin{equation}
    P_\psi= \frac{1}{2}\dot{\psi}^2-U(\psi), \quad \rho_\psi= \frac{1}{2}\dot{\psi}^2+U(\psi).
\end{equation}
The scalar field potential $U(\psi)$ es depicted in Fig.(\ref{fig:Potential-2}).
The global minimum of $U(\psi)$ is $U_T$, but there is also a local minimum $U_F$ (the false vacuum).

We have considered that the early universe is dominated by two fluids because in our scheme of EU scenario, during the static regime, the scalar field $\psi$ remains static at the false vacuum, in contrast to standard EU models where the scalar field rolls on the asymptotically flat part of the scalar potential \cite{delCampo:2009kp}. Then, in order to obtain a static universe under these conditions, we have to included a standard perfect fluid. For simplicity we will consider that there are no interactions between the standard perfect fluid and the scalar field.
The Friedmann-Raychaudhuri equations become

\begin{equation} \label{ecampo1}
H^2+\frac{1}{a^2}+H\frac{\dot{\phi}}{\phi}=\frac{\rho}{3\phi}+
\frac{\omega}{6}\left(\frac{\dot{\phi}}{\phi}\right)^2+\frac{V}{3\phi},
\end{equation}

\begin{equation}\label{ecampo2}
2\frac{\ddot{a}}{a}+H^2+\frac{1}{a^2}+
\frac{\ddot{\phi}}{\phi}+2H\frac{\dot{\phi}}{\phi} +
\frac{\omega}{2}\left(\frac{\dot{\phi}}{\phi}\right)^2
-\frac{V}{\phi} =-\frac{P}{\phi}.
\end{equation}
The field equation for the JBD field is
\begin{center}
    \begin{equation}\label{ecampo3}
    \ddot{\phi}+3H\dot{\phi}=\frac{\rho-3P}{2\omega+3}+\frac{2}{2\omega+3}
    \left[2V-\phi V' \right],
    \end{equation}
\end{center}
where $V' = dV(\phi)/d\phi$, $\rho= \rho_f + \rho_\psi $, $P=P_f+P_\psi$ and dots represent derivatives with respect to cosmological time.

The conservation equations for the scalar field and perfect standard fluid are
\begin{equation}\label{conser1}
\ddot{\psi}+3H \dot{\psi} = -\dfrac{\partial U (\psi)}{\partial \psi},
\end{equation}
and
\begin{equation}\label{conser2}
\dot{\rho_f} +3 H \left(\rho_f+P_f\right)=0,
\end{equation}
respectively.\\
The static universe is characterized by the conditions
$a=a_0=\text{Const}.$, $\dot{a}_0=\ddot{a}_0=0$ and
$\phi=\phi_0=\text{Const}.$, $\dot{\phi}_0=\ddot{\phi}_0=0$,
$\psi=\psi_F$. From Equations \eqref{ecampo1}, \eqref{ecampo2} and
\eqref{ecampo3} the static solution for a universe dominated by a
scalar field placed in a false vacuum and a standard perfect fluid
is obtained if the following conditions are satisfied:
\begin{equation}\label{cond1}
    a_0^2=\frac{3}{V'_0},
\end{equation}
\begin{equation}\label{cond2}
    \rho_{f0}=-U_F-V_0+V'_0 \phi_0,
\end{equation}
and
\begin{equation}\label{cond3}
    \gamma=2\frac{\phi_0}{a_0^2 \:\rho_{f0}},
\end{equation}
where $V_0=V(\phi_0)$ and
$V'_0=\left(dV(\phi)/d\phi\right)_{\phi=\phi_0}$ and $\rho_{f0}$ is
the energy density of the perfect fluid present in the static
universe. These equations connect the equilibrium values of the
scale factor and the JBD field with the energy density and the JBD
potential at the equilibrium point. We now study the stability of
this solution against small homogeneous and isotropic perturbations.
In order to do this, we consider small perturbations around the
static solutions for the scale factor, JBD field and inflaton field.

We set
\begin{equation}\label{pert1}
    a(t)=a_0 \left[1+\alpha(t) \right],
\end{equation}
\begin{equation}\label{pert2}
    \phi(t)=\phi_0 \left[1+\beta(t) \right],
\end{equation}
\begin{equation}\label{ndeltapsi}
\psi(t) = \psi_F[1+ \lambda(t)].
\end{equation}
Then, we have
\begin{equation}\label{pert3}
\rho_f(t) = \rho_{f0}\left(\frac{a_0}{a(t)}\right)^{3\gamma} =
\rho_{f0}\left(\frac{a_0}{a_0(1 + \alpha(t))}\right)^{3\gamma}
\approx \rho_{f0}\Big(1 - 3\gamma \,\alpha(t)\Big) = \rho_{f0} + \delta \rho_{f0}\,,
\end{equation}
where $\alpha(t) \ll 1$, $\beta(t) \ll 1$ and $\lambda (t) \ll 1$
are small perturbations. On the other hand, we have that at linear
order in $\lambda$
\begin{eqnarray}\label{ndeltar2}
 \delta \rho_\psi = 2\dot{\psi}_F\,\delta \dot{\psi} +
U'_F\,\delta \psi\,,\\
\delta P_\psi = 2\dot{\psi}_F\,\delta \dot{\psi} - U'_F\,\delta
\psi\,,\label{ndeltapsi2}
\end{eqnarray}
where $\delta \psi = \psi_F\,\lambda(t)$ and $\delta \dot{\psi} =
\psi_0\,\dot{\lambda}(t)$.
As in our case (inflaton field in a local minimum of its potential)
$\dot{\psi}_F=0$ and $U'_F=0$ then we obtain that $\delta \rho_\psi
= \delta P_\psi = 0$, at linear order.
Then, we have that the evolution of $\lambda$ is determined only by
the following equation obtained from Eq.~(\ref{conser1}) and
Eq.~(\ref{ndeltapsi})
\begin{equation}\label{nlambda} \ddot{\lambda} + U''_F\,\lambda = 0\,,
\end{equation}
where $U''_F = \left(d^2U(\psi)/d\psi^2\right)_{\psi=\psi_F}$. We
can note that the evolution of $\lambda$ is completely decoupled
from the evolution of $\alpha$ and $\beta$ and correspond to an
oscillatory behavior given that $U''_F >0$ in our case.

By introducing the expressions (\ref{pert1}), (\ref{pert2}) and
(\ref{pert3}) into equations (\ref{ecampo2}) and (\ref{ecampo3}),
and keeping terms at the linear order in $\alpha$, $\beta$ and
$\lambda$, we obtain the following set of coupled equations
\begin{equation}\label{acop1}
\ddot{\alpha} -\left[\frac{1}{a_0^2} +3\frac{\left(\gamma-1\right)}{a_0^2}\right]\alpha
+\frac{\ddot{\beta}}{2}-\frac{\beta}{a_0^2} = 0,
\end{equation}
and
\begin{equation}\label{acop2}
    \left(3+2\omega\right) \ddot{\beta} -\left(\frac{6}{a_0^2}-2\phi_0 V''_0\right) \beta +\left(4-3\gamma\right)\frac{6}{a_0^2} \alpha =0,
\end{equation}
where $V''_0=\left(d^2V(\phi)/d\phi^2\right)_{\phi=\phi_0} $.\\
From the system of equations \eqref{acop1} and \eqref{acop2} we can obtain the frequencies for small oscillations
\begin{multline}
    w^{2}_{\pm} = \frac{1}{a_0^2 \left(3+2\omega\right)} \left[a_0^2 \phi_0 V''_0 -6 +\omega\left(2-3\gamma\right)\right. \\
    \pm \left.\sqrt{\left[-6+a_0^2 \phi_0 V''_0+2\omega-3\omega\gamma\right]^2 + 2\left(3+2\omega\right)\left(-6+a_0^2 \phi_0 V''_0\left[3\gamma-2\right]\right)} \right].
\end{multline}
Note that the static solution is stable if the inequality $ w_{\pm}^2 > 0 $ is satisfied.
\\ Then assuming that the parameter $\omega$ satisfies the constraint, $ \left(3 + 2 \omega \right)> 0 $, we find that the following inequalities must be fulfilled in order to have a stable static solution
\begin{equation}\label{condgamma}
    \frac{2}{3} <\gamma<\frac{4}{3},\quad \gamma>\frac{4}{3}
\end{equation}
\begin{equation}\label{condomega}
    -\frac{3}{2}<\omega<-18\frac{\left(\gamma-1\right)}{\left(2-3\gamma\right)^2},
\end{equation}
and
\begin{equation}\label{vdosprima}
    2\left(6+\omega\right)-3\left(3+\omega\right)\gamma+\sqrt{3}\left|4-3\gamma\right|\sqrt{3+2\omega}<a_0^2 \phi_0 V''_0<\frac{6}{3\gamma-2}.
\end{equation}
From these inequalities we can conclude that for a universe dominated by a scalar field in a false vacuum and a perfect standard fluid, it is possible to find a solution where the universe is static and stable.

In particular, we study numerical solutions to the equations \eqref{ecampo2}, \eqref{ecampo3}, \eqref{conser1} and \eqref{conser2}, for initial conditions close to the static solution.
We address the case in which the parameters take the following values that satisfy the stability conditions \eqref{condgamma}, \eqref{condomega} and \eqref{vdosprima}
\begin {equation}
\gamma = 1; \quad \phi_0 = 0.5; \quad a_0 = 1; \quad U_F = 1; \quad
\psi_F = 1,
\end {equation}
where we have used units in which $8\pi G = 1$. From the conditions
of the static solution \eqref{cond1}, \eqref{cond2}, \eqref{cond3}
and the stability condition \eqref{vdosprima} we obtain the values
for $ V_0 $ , $ V'_0 $ and $ V''_0 $.

\begin{figure}[t]
    \centering
    \begin{tabular} {llll}
       \includegraphics [scale=0.6] {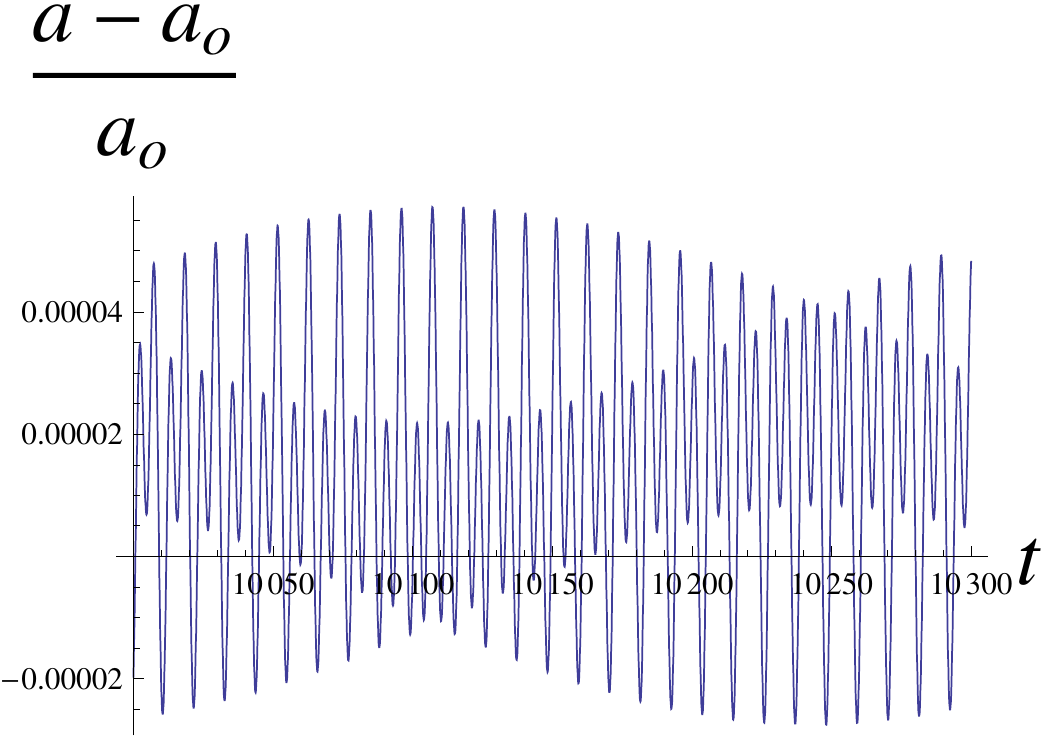}
        & \includegraphics [scale=0.6] {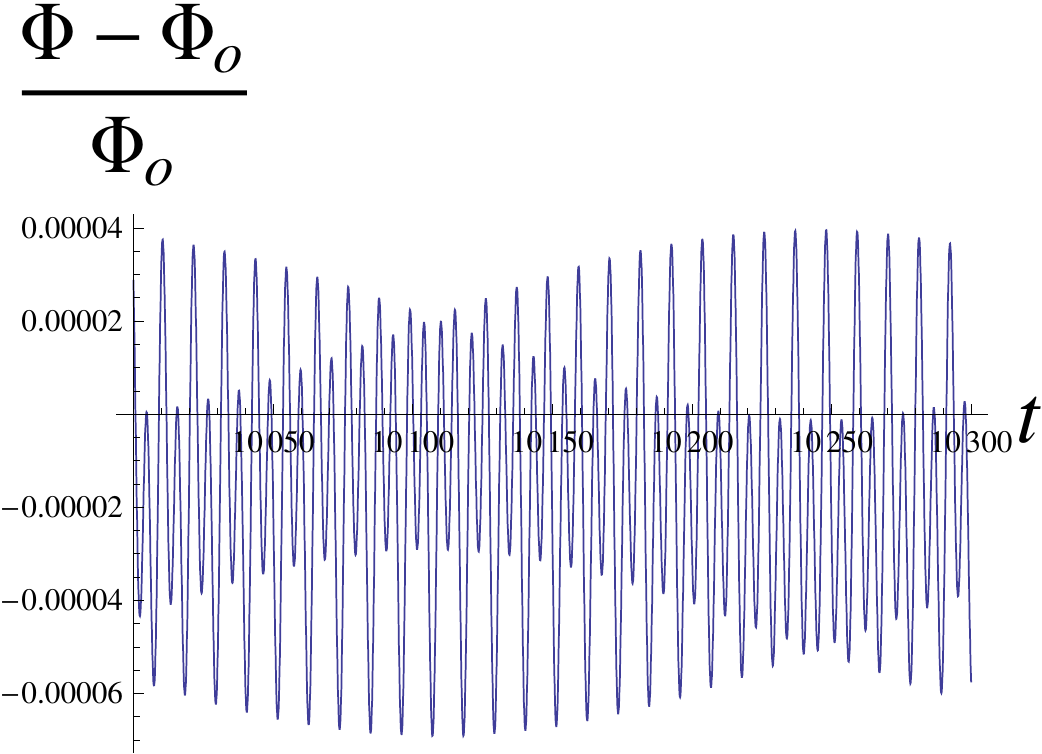} \\
        \includegraphics [scale=0.6] {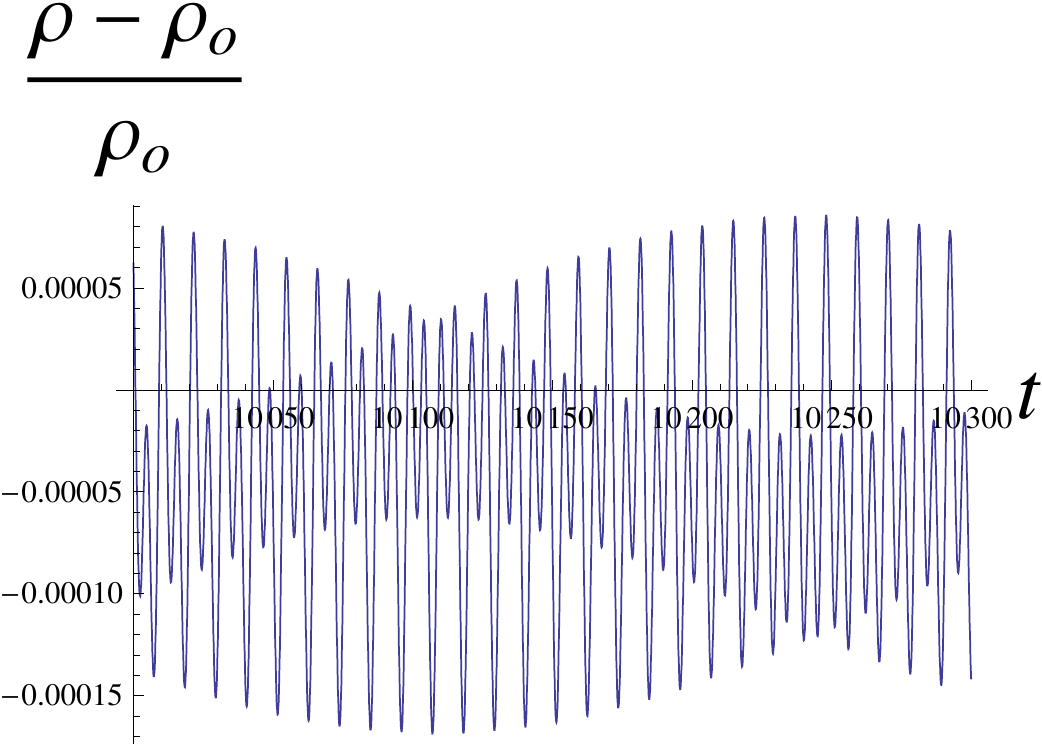}
        & \includegraphics [scale=0.6] {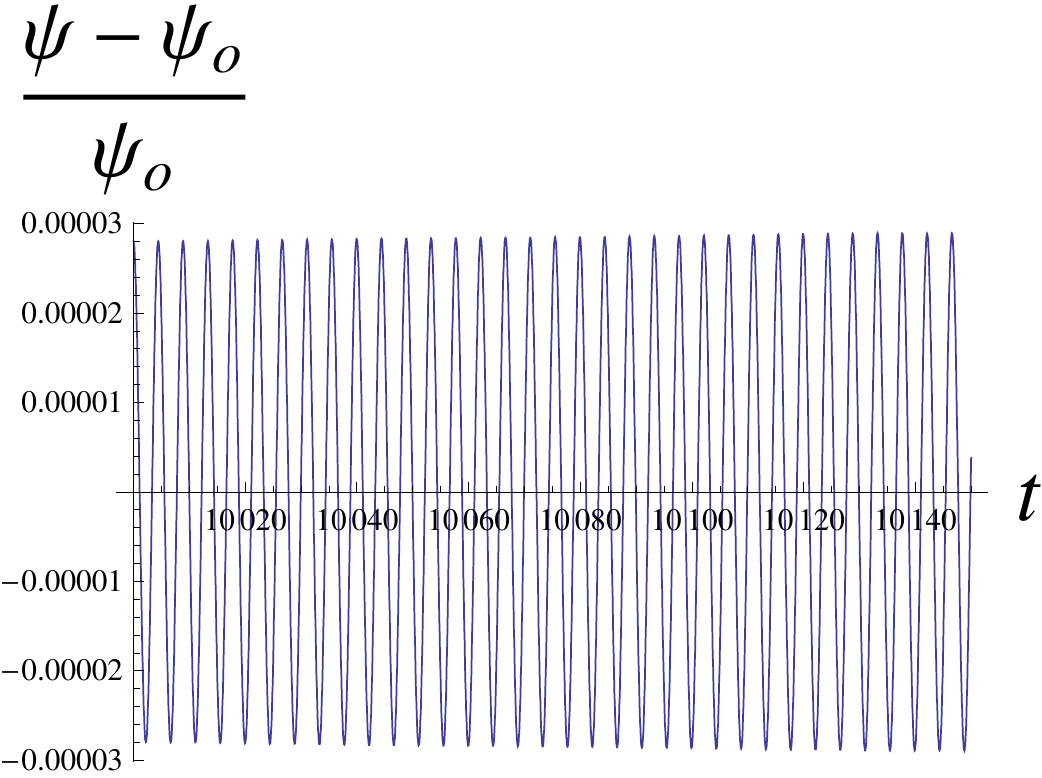}
   \end{tabular}
\caption {Behavior of the scale factor $a$, the JBD field $\phi$,
the perfect fluid density $\rho$ and the inflaton field $\psi$ as a
function of cosmological time.} \label{fig:evolucion}
\end{figure}

In particular, for the numerical solution, we consider the following
inflaton potential
\begin{equation}
U(\psi) = U_F + \left(\psi - \psi_F\right)^2 \,,
\end{equation}
where we have used  that around the local equilibrium point $\psi
\sim \psi_F$ the inflaton potential potential can be approximated to
a quadratic function.

Also, we consider the following JBD potential which satisfies the
static and stability conditions discussed above
\begin{equation}
 V (\phi) = V_0 + V'_0 \left(\phi-\phi_0\right)+ \frac{V''_0}{2} \left(\phi-\phi_0\right)^2.
\end{equation}

In Fig.~(\ref{fig:evolucion}) it is shown one numerical solution
corresponding to a universe starting from an initial state not in
the static solution but close to it. We can note that while the
scalar field makes small oscillations around the false vacuum, the
scale factor, JBD field and perfect fluid density present small
oscillations around their equilibrium values. This tells us that the
static solution is classically stable.

Then, in a purely classical field theory if the universe is static
and supported by the scalar field located at the false vacuum $U_F$,
the universe remains static forever. Quantum mechanics makes things
different because the scalar field $\psi$ can tunnel through the
barrier and by this process create a small bubble where the field
value is $\psi_T$. Depending on the background where the bubble
materializes the bubble could expand or collapse, see
Refs.~\cite{Labrana:2011np, Simon:2009nb, Fischler:2007sz}.

\section{Bubble Nucleation}\label{sec:tunel}

In this section we study the tunneling process of the scalar field
$\psi$ from the false vacuum $U_F$ to the true vacuum $U_T$, in the
potential $U(\psi)$ shown in Fig.~(\ref{fig:Potential-3}) and the
consequent creation of a bubble of true vacuum in the background of
an Einstein static universe, in the context of a JBD Theory.
In particular, we will consider the nucleation of a spherical bubble
of true vacuum  within the false vacuum. We will assume that the
layer which separates the two phases (the wall) has a negligible
thickness compared to the size of the bubble (the usual thin-wall
approximation). The energy budget of the bubble consists of latent
heat (the difference between the energy densities of the two phases)
and surface tension.

In order to study the tunneling process in this model, we have to
deal with the non-standard gravitational interaction of the JBD
theory.
This was done in Ref.~\cite{Holman:1989gh} and here we reproduce and
adapt these results to the EU scheme. Regarding the study of bubble nucleation in JBD theories in other contexts see Refs.~\cite{Lee:2008hz, Kim:2010yr}.

We begin with the JBD action (\ref{accion}) described in
Sect.~\ref{sec:estadoestatico} and perform a weyl rescaling in order
to remove the non-standard coupling of $\phi$ to $\cal R $.

\begin{equation}\label{trans}
    g_{\mu\nu} = \Omega^{-2} (x) \tilde{g}_{\mu\nu},
\end{equation}
\begin{equation}\label{fitilde}
    \tilde{\phi} =  \sqrt{\frac{2\omega+3}{8\pi G}}  \ln \left(\phi / \phi_G\right),
\end{equation}
where $\phi_G^{-1} = G $, here $G$ is the value of the Newton
constant observed in the present. We choose the conformal factor
$\Omega(x)$ to be $\Omega = \sqrt{8\pi G \phi} $. Then, the action
in the rescaled theory can be expressed as follow
\begin{multline}\label{rescalada}
S[\tilde{g},\tilde{\phi},\psi]= \int d^4x \sqrt{-\tilde{g}}
\left[\frac{1}{16\pi G} \tilde{{\cal R}} - \frac{1}{2}
\tilde{g}^{\mu\nu} \tilde{\nabla}_\mu \tilde{\phi}
\tilde{\nabla}_\nu \tilde{\phi} +
    W(\tilde{\phi})\right. %\\
    \left. + \frac{b}{2} \tilde{g}^{\mu\nu} \nabla_\mu \psi \nabla_\nu
    \psi - b^2 U(\psi) + b^2\,{\cal L}_m\right],
\end{multline}

where we have defined
\begin{equation}
    b\equiv\exp{\left(-4\sqrt{\frac{\pi G}{2\omega+3}}\tilde{\phi}\right)}\,,
\end{equation}

\begin{equation}
W(\tilde{\phi}) = \dfrac{V(\phi(\tilde{\phi}))}{(8\pi)^2 G^2
\phi^2}=\frac{V(\phi(\tilde{\phi}))}{(8\pi)^2}\exp
{\left(-4\sqrt{\frac{2\pi G}{2\omega+3}}\tilde{\phi}\right)}.
\end{equation}

In the rescaled theory the probability of nucleation per unit of
physical volume per time is given by
\begin{equation}
    \bar{\Gamma} (\bar{t}) \equiv \dfrac{dP (\bar{t})}{d \bar{V} (\bar{t}) d\bar{t}} \:,
\end{equation}
where $P (\bar{t})$ is  the bubble probability of nucleation and
$\bar{V} (\bar{t})$ is the  physical volume measured in the rescaled
system at time $\bar{t}$. Therefore, the equivalent rate in the
original theory is given by
\begin{equation}\label{taza}
    \Gamma(t) \equiv \frac{dP (t)}{dV(t) dt} = \dfrac{dP
        (\bar{t})}{\left(\Omega^3 d \bar{V}\right)\left( \Omega
        d\bar{t}\right)} = \Omega^{-4} (t) \bar{\Gamma} (\bar{t}),
\end{equation}
where we have used that $P(t)=P\left[\bar{t}(t)\right]$, see
Ref.~\cite{Holman:1989gh}.

In the rescaled formulation of the theory, the gravitational
interaction has the standard form, and so we can expect
gravitational effects similar to those of the standard theory.

Nevertheless as it is usually, in order to eliminate the problem of  predicting the reaction of the geometry to an essentially a-causal quantum jump (the materialization of the bubble), we neglect during
this computation the gravitational back-reaction of the bubble onto
the space-time geometry.
The gravitational back-reaction of the bubble will be consider in
the next Section when we study the evolution of the bubble after its
materialization.
Thus, in this approach only the matter field $\psi$ remains as a dynamical quantity in the action (\ref{rescalada}).
Then, by following Ref.\cite{Holman:1989gh} we determinate the rate of nucleation of a bubble of true vacuum in the remaining dynamical theory for the field $\psi$.  The nucleation rate $ \bar {\Gamma} $ is given by
\begin{equation}
    \bar{\Gamma}(t)=A \exp{\left(-S_E\right)}
\end{equation}
where $S_E$ is the Euclidean action.

It is shown in Ref.~\cite{Holman:1989gh} that the coefficient $A$ has a
net factor of $b^2$ with respect to a theory in which $b = 1$,
this entails that $\bar{\Gamma}(t)=b^2 \Gamma_0$, where $ \Gamma_0 $
is the nucleation rate for a normal scalar field theory with
potential $ U (\psi) $. Since $ b $ is a function of the
time-dependent JBD field, then we have a time-dependent nucleation
rate in the rescaled theory. However, this time dependence
disappears if we return to the original theory. The equations
\eqref{trans} and \eqref{fitilde}, together with the definition of
$\phi_G $ and $\Omega$ imply that $ b = \Omega^2 $. Using the
equation \eqref{taza} we find that the nucleation rate of the
original theory is, see \cite{Holman:1989gh}
\begin{equation}
    \Gamma(t) =\Omega^{-4} (t) \bar{\Gamma}
    (\bar{t})=\left(b^{-2}\right)\left(b^2 \Gamma_0\right)=\Gamma_0 \,.
\end{equation}

The computation of $\Gamma_0$ in the context of the EU scheme, that is, the tunneling process of the scalar field $\psi$ in a background of Einstein static universe in the context of General Relativity, was developed in Refs.~\cite{Labrana:2011np, Labrana:2013kqa, Labrana:2014yta}, where it was found that

\begin{equation}\label{corrective}
    \Gamma_0 \approx \exp \left[-\frac{27 \sigma^4 \pi}{2 \varepsilon^3}
    \left(1-\frac{9\,\sigma^2}{2\,  \varepsilon^3 \,a^2_0 }\right)\right],
\end{equation}
where $\varepsilon$ is the difference of energy density between the
two phases (latent heat) and $\sigma $ is the energy density of the wall.

\section{Classical Evolution of the Bubble on Jordan-Brans-Dicke Theories}\label{sec:evolucionbur}

In this section we study the evolution of true vacuum bubble
after its nucleation via quantum tunnel effect.
During this study we are going to consider the gravitational
back-reaction of the bubble.

We follow the approach and notation used in Ref.~\cite{Berezin:1987bc}
where it is assumed that the bubble wall separates spacetime into
two parts. The bubble wall is a timelike, spherically symmetric
hypersurface, the interior of the bubble is our universe, according
to the extended open inflation scheme
\cite{linde, re8, delC1, delC2, Balart:2007je}
and the exterior correspond to the static universe discussed in Sec.
II.
In particular, we will use the scheme developed in Ref.~\cite{Sakai:1992ud} regarding the Darmois-Israel junction conditions \cite{Israel:1966rt, darmois} applied to a JBD Theory.

Let us start by summarize the formalism developed by Berezin, Kuzmin and Tkachev in Ref.~\cite{Berezin:1987bc} for the analysis of a thin wall bubble in the context of General Relativity before to study this problem in the context of a Jordan-Brans-Dicke-Theory.

We begin by defining a time-like spherically symmetric hypersurface $\Sigma $, representing the world surface of the wall dividing space-time into two regions, $ V^{+} $ (outside) and $ V^{-} $ (inside). Also, we define a space-like vector $ N_{\mu} $, which is orthogonal to $ \Sigma $ and points from $ V^{-} $ to $ V^{+} $. It is convenient to introduce a normal Gaussian coordinate system $ \left (n, x^i \right) $ such that the hypersurface $ n = 0 $ corresponds to $ \Sigma $. If we assume that the wall is infinitely thin, its surface energy-momentum tensor is
\begin{equation} \label {dosuno}
S_{ij} \equiv \lim_{\delta \to 0} \int_{-\delta}^{\delta}  \! T_{i j} \, dn.
\end{equation}

Using the extrinsic curvature tensor defined by $ K_{ij} \equiv N_{i;j}$ and the Einstein's equations we can express the joint conditions on the wall as follows, see \cite{Israel:1966rt}
\begin{equation} \label{dosdos}
\left[K_{ij}\right]^{\pm} =-\kappa^{2} \left(S_{ij} -\frac{1}{2} h_{ij} Tr \, S \right),
\end{equation}

\begin{equation} \label{dostres}
-S_{i|j}^{\,j} = \left[T_{i}^{\,n} \right]^{\pm},
\end{equation}

\begin{equation} \label{doscuatro}
\frac{K_{j}^{\,i \, +} + K_{j}^{\,i \, -}}{2} S_{i}^{j} = \left[T_{n}^{\,n} \right]^{\pm},
\end{equation}
where $\kappa^{2} \equiv 8 \pi G $, $h_{ij}$ is the 3-metric on $ \Sigma $, and $ |$ denotes the three-dimensional covariant derivative. We use the brackets to represent the difference between the outer and inner values of any field variable, for example: $\left[F\right]^{\pm} \equiv F^{+}-F^{-}$. By using Eq.\eqref{dosdos} and Eq.\eqref{doscuatro} we eliminate $K_{j}^{\,i \, -}$ obtaining:
\begin{equation} \label{doscinco}
K_{j}^{\,i \, +}  S_{i}^{j} + \frac{\kappa^{2}}{2} \left\{  S_{j}^{i} S_{i}^{j}- \frac{1}{2} \left(Tr \, S\right)^2 \right\} = \left[T_{n}^{\,n} \right]^{\pm}.
\end{equation}
Since Eq.\eqref{dostres} and Eq.\eqref{doscinco} do not contain $K_{j}^{\,i \, -}$, they may be useful when the geometry of interior of the bubble is unknown.

We are going to assume that the bubble wall $\Sigma$ is spherically symmetric, then the intrinsic metric on the shell $\Sigma$ is given by
\begin{equation}\label{dosseis}
\left. ds^2\right|_{\Sigma} = d\tau^2 - R^2(\tau)\left(d\theta^2+\sin^2\! \theta \, d\varphi^2 \right),
\end{equation}
where $ R (\tau) $ is the radius of circumference of the wall and $ \tau $ the proper time of the wall.
The surface energy-momentum tensor $S_{ij}$ can be written as a perfect fluid:
\begin{equation} \label{dossiete}
S_{ij} = \left(\sigma-\tilde{\omega}\right)u_{i}u_{j}+\tilde{\omega} h_{ij},
\end{equation}
where $ \sigma $, $\tilde{\omega}$ and $u^i=(1,0,0)$  are the surface energy density, surface pressure and a unit time-like vector tangent to $ \Sigma $, respectively. Then we can rewrite Eq.\eqref{dostres} and Eq.\eqref{doscinco} as follow
\begin{equation} \label{dosocho}
\dfrac{d \sigma}{d \tau} + 2 \dfrac{d R}{d \tau} \frac{\left(\sigma+\tilde{\omega}\right)}{R} = \left[T_{\tau}^{\,n} \right]^{\pm},
\end{equation}
\begin{equation} \label{dosnueve}
-\sigma K_{\tau}^{\,\tau \, +} +2 \tilde{\omega} K_{\theta}^{\,\theta \, +} + \kappa^{2} \sigma \left(\frac{\sigma}{4} + \tilde{\omega} \right) = \left[T_{n}^{\,n} \right]^{\pm}.
\end{equation}
As was discussed in Ref.\cite{Berezin:1987bc}, Eq.\eqref{dosocho} and Eq.\eqref{dosnueve} determine the evolution of $\sigma$ and $R$, in the context of General Relativity, as they give us information on how its radius evolves and about the energy density that accumulates on the outer side of the wall.

In our case, in order to derive the equations of motion of the bubble wall in the Jordan-Brans-Dicke theory we are going to follow a scheme similar to that discussed above, see Ref.\cite{Sakai:1992ud}.

From the JBD action Eq.~(\ref{accion}) we obtained the following field equations

\begin{equation} \label{trestres}
{\cal R}_{\mu\nu} - \frac{1}{2} g_{\mu\nu} {\cal R} = \frac{1}{\phi} \tilde{T}_{\mu \nu},
\end{equation}
\begin{equation} \label{trescuatro}
\displaystyle \Box \phi = \frac{1}{2\omega+3} \left[Tr \textit{T} - 2\phi V'(\phi) + 4 V(\phi)\right],
\end{equation}
where
\begin{equation} \label{trescinco}
\tilde{T}_{\mu\nu} \equiv  \frac{\omega}{\phi} \nabla_\mu \phi \nabla_\nu \phi - \frac{\omega}{2\phi} g_{\mu\nu} \left(\nabla \phi\right)^2 +\nabla_\mu \nabla_\nu \phi- g_{\mu\nu} \Box\phi + V(\phi) g_{\mu\nu} + T_{\mu\nu},
\end{equation}
and primes denote derivatives with respect to $\phi$.
In this context it is useful to introduce the following surface energy-momentum tensor defined over $ \Sigma $
\begin{equation} \label{tressiete}
\tilde{S}_{ij} \equiv  \lim_{\epsilon \to 0} \int_{-\epsilon}^{\epsilon}  \! \tilde{T}_{i j} \, dn = S_{ij} -\frac{1}{2} \left(\frac{1}{2\omega+3}\right)h_{ij} Tr \textit{S} ,
\end{equation}
where the second equality is obtained from Eq.\eqref{trescuatro} and Eq.~\eqref{trescinco}. With these definitions we can use the formalism described above (for General Relativity) in order to study the evolution of the bubble in the Jordan-Brans-Dicke context, we only have to replace $T_{\mu \nu}$, $S_{ij}$ and $\kappa^2$ by $\tilde{T_{\mu \nu}}$, $\tilde{S_{ij}}$ y $1/\phi$, respectively. The junction condition for the Jordan-Brans-Dicke field $ \phi $ is obtained from Eq\eqref{trescuatro}, see Ref.\cite{suffern}, in particular we obtain:
\begin{equation} \label{tresocho}
\phi^{+} = \phi^{-}, \: \: \: \: \: \: \left[\phi_{,n}\right]^{\pm} = \frac{1}{2\omega+3} Tr \textit{S}.
\end{equation}
We note from this condition that in general  $V^{+}$ and $V^{-}$ can not be both homogeneous.

Now we rewrite Eq.~\eqref{dosdos}, Eq.~\eqref{dostres} and Eq.~\eqref{doscinco} in the context of a Jordan-Brans-Dicke theory
\begin{equation}
\left[K_{ij}\right]^{\pm}= -\frac{1}{\phi} \left(\tilde{S}_{ij}-\frac{1}{2} h_{ij} Tr \tilde{S}\right),
\end{equation}
\begin{equation}
-\left\{\frac{\tilde{S}_{i}^{j}}{\phi}\right\}_{|j} = \frac{\left[\tilde{T}_{i}^{n}\right]^{\pm}}{\phi},
\end{equation}
\begin{equation}
K_{j}^{i \, +} \tilde{S}_{i}^{j} + \frac{1}{2\phi} \left\{\tilde{S}_{j}^{i} \tilde{S}_{i}^{j} -\frac{1}{2} \left(Tr \tilde{S}\right)^2 \right\} = \left[\tilde{T}_{n}^{n}\right]^{\pm}.
\end{equation}
By using the equations \eqref{trescinco} and \eqref{tressiete}, we obtain
\begin{equation}
\left[K_{ij}\right]^{\pm}= -\frac{1}{\phi} \left(S_{ij}-\frac{1}{2} h_{ij}\left\{1-\frac{1}{2\omega+3} \right\} Tr \textit{S}\right),
\end{equation}
\begin{equation} \label{trestrece}
-S_{i|j}^{j} =\left[ T_{i}^{n} \right]^{\pm} ,
\end{equation}
\begin{equation} \label{trescatorce}
K_{j}^{i \, +} S_{i}^{j} + \frac{1}{2\phi} \left(S_{j}^{i} S_{i}^{j} -\frac{1}{2} \left(Tr S\right)^2 \left\{1-\frac{1}{2\omega+3} \right\} \right) = \left[T_{n}^{n}\right]^{\pm}.
\end{equation}
As was mentioned, the exterior of the bubble ($V^{+}$) is described by the metric of a closed Friedmann-Robertson-Walker universe. We write this metric as follows:
\begin{equation} \label{tres15}
ds^2 = g^{+}_{\mu\nu} dx_{+}^{\,\mu} dx_{+}^{\,\nu} = dt_{+}^{2} - a^2(t_{+}) \left\{d\chi_{+}^2 + r^2 (\chi_{+})\left(d\theta^2+ \sin^2\! \theta \,d\varphi^2 \right)\right\},
\end{equation}
where $r(\chi_{+}) = \sin (\chi_{+}) $.

We can note from the conditions \eqref{tresocho} that if one of the regions of space-time is homogeneous the other region is not. As in our case the outer region ($V^{+}$) is a homogeneous  ES universe, then the region $V^{-}$ is generally inhomogeneous.

Substituting Eq.~\eqref{dossiete} into Eq.~\eqref{trestrece} and Eq.~\eqref{trescatorce}, we get
\begin{equation} \label{tres16}
-\sigma K_{\tau}^{\,\tau \, +} +2 \tilde{\omega} K_{\theta}^{\,\theta \, +} + \frac{1}{4\phi} \sigma \left(\sigma + 4\tilde{\omega} \right) +\frac{1}{2\omega+3} (-\sigma+2\tilde{\omega})^2 = \left[\tilde{T}_{n}^{\,n} \right]^{\pm},
\end{equation}
\begin{equation} \label{tres17}
\dfrac{d\sigma}{d\tau}+\dfrac{dR}{d\tau} \,\frac{\left(\sigma+\tilde{\omega}\right)}{R} = \left[T_{\tau}^{\,n}\right]^{\pm}.
\end{equation}
The conditions of continuity for the metric on $\Sigma$ are obtained from \eqref{dosseis} and \eqref{tres15} as
\begin{equation} \label{tres18}
R(\tau) =a(t_{+}(\tau)) r(\chi_{+}(\tau))|_{\Sigma}  , \:\:\: \quad d\tau^2=dt_{+}^2-a^2 d\chi_{+}^2|_{\Sigma}.
\end{equation}
As usual, we consider that the content of matter in the outer region $V^{+}$  and in the inner region $V^{-}$ is a perfect fluid, with the following energy-momentum tensor
\begin{equation}
T^{\pm}_{\mu\nu} = \left(p^{\pm}+\rho^{\pm} \right)U^{\pm}_{\mu} U^{\pm}_{\nu} +p^{\pm} g^{\pm}_{\mu\nu},
\end{equation}
where $ p $, $ \rho $ and $U_{\mu}$ are the pressure, energy density and 4-velocity of the perfect fluid, respectively. Then, by following Ref.~\cite{Sakai:1992ud}, we can rewrite Eq.\eqref{tres16} and Eq.\eqref{tres17} as
\begin{multline} \label{primera}
\dfrac{d\left(\beta_+ v_+\right)}{dt_+} = -\beta_+\left\{\left(1-2\frac{\tilde{\omega}}{\sigma}\right) v_+ H-\frac{2}{R} \dfrac{dr}{d\chi_+}\frac{\tilde{\omega}}{\sigma}\right\} \\+\frac{1}{4\phi}\left\{\sigma+4\tilde{\omega}+\frac{1}{2\omega+3} \frac{\left(-\sigma+2\tilde{\omega}\right)^2}{\sigma}\right\}-\frac{\left[\beta^2\left(v^2 \rho + p\right)\right]^{\pm}}{\sigma},
\end{multline}
\begin{equation}\label{segunda}
\dfrac{d \sigma}{d t_+} = -2 \dfrac{d R}{d t_+} \frac{\left(\sigma+\tilde{\omega}\right)}{R} + \left[\beta v \left(\rho + p\right)\right]^{\pm},
\end{equation}
\begin{equation} \label{tercera}
\dfrac{d R}{d t_+} = \dfrac{dr}{d\chi_+} v_+ + H R,
\end{equation}
where
\begin{equation}
v_+\equiv a \dfrac{d \chi_+}{d t} , \quad \beta_+ \equiv \dfrac{d t_+}{d \tau} \equiv \frac{1}{\sqrt{1-v_+^2}} , \quad H \equiv \dfrac{d a/d t_+}{a}.
\end{equation}
So far we have summarized the calculations of Ref.~\cite{Sakai:1992ud} related to the evolution of a bubble in a Jordan-Brans-Dicke theory in a general background. Now we will concentrate on our particular case, where the bubble materializes and then evolves in a background corresponding to a closed and static universe, as the one described in Sec.~\ref{sec:estadoestatico}.

We assume that the outer region of the bubble is static, that is
\begin{equation}\label{condiciones}
     a\left(t\right)=a_0,\:\:\: \phi\left(t\right) = \phi_0,
\end{equation}
with $ a_0 $ and  $ \phi_0 $ constants, defined in Sec.\ref{sec:estadoestatico}.

In the outer region, as content of matter we consider a scalar field $\psi$ localized in a false vacuum, and a perfect fluid, see discussion in Sec.~\ref{sec:estadoestatico}. Then we have
\begin{equation}
\rho_{+}=\rho_{f0} + U_F.
\end{equation}
\begin{equation}\label{baroextint}
    p_{+}=\left(\gamma_{f}-1\right)\rho_{f0}-U_F.
\end{equation}
In the inner region, the matter content is described by the scalar field in the true vacuum ($\gamma_{-}=0)$. Then, in the inner region we have
\begin{equation}
p_{-}=-\rho_{-} = - U_T.
\end{equation}
Now, we develop the derivative of the left hand side of the equation (\ref{primera})
\begin{equation}
\displaystyle \dfrac{d\left(\frac{v_+}{\sqrt{1-v_+^2}} \right)}{dt_+} = \frac{v_+^2 \: \dfrac{d v_+}{dt_+}}{\left(1-v_+^2\right)^{3/2}} + \frac{\dfrac{d v_+}{dt_+}}{\left(1-v_+^2\right)^{1/2}} = \frac{\dfrac{d v_+}{dt_+}}{\left(1-v_+^2\right)^{3/2}}.
\end{equation}
Then the equation \eqref{primera} takes the form
\begin{multline}\label{final2}
\frac{\dfrac{d v_+}{dt_+}}{\left(1-v_+^2\right)^{3/2}}  = \frac{1}{\sqrt{1-v_+^2}}\left\{\frac{2}{R} \left(\tilde{\gamma}-1\right)\dfrac{dr}{d\chi_+}\right\} +\frac{1}{4\phi}\left\{4\tilde{\gamma} \sigma - 3\sigma + \frac{\sigma}{2\omega+3} \left(-3+2\tilde{\gamma}\right)^2 \right\}\\
 +\frac{\varepsilon}{\sigma}+\frac{\gamma_+ \rho_+}{\left(1-v_+^2\right)\sigma},
\end{multline}
where we have defined $\varepsilon = (\rho_{+} - \rho_{-})/\sigma$.

By using $r\!\left(\chi_+\right) = \sin \!\left(\chi_+\right)$ and $R = a_0 r$, we obtain
\begin{equation}
 \dfrac{d\:r\left(\chi_+\right)}{d\chi_+} = \cos \!\left(\chi_+\right) =\sqrt{1- \sin^2\! \left(\chi_+\right)} = \sqrt{1-\left(R/a_0\right)^2}.
\end{equation}
If we introduce this into the Eq.~\eqref{tercera}, we have
\begin{equation}\label{R1}
\dfrac{d R}{d t_+} = \sqrt{1-\left(R/a_0\right)^2} \,v_+.
\end{equation}

We are going to consider that the matter content of the bubble wall is the same as that of the outer region ($V^+$), then we have $\tilde{\gamma}=\gamma_f$.

Therefore, we write Eq.~\eqref{final2} as follows
\begin{multline}\label{V1}
\frac{\dfrac{d v_+}{dt_+}}{\left(1-v_+^2\right)^{3/2}}  = \frac{1}{\sqrt{1-v_+^2}}\left\{\frac{2}{R} \left(\gamma_f-1\right)\left(\sqrt{1-\left(R/a_0\right)^2}\right)\right\} \\+\frac{1}{4 \phi}\left\{4\,\gamma_f\, \sigma - 3\sigma  + \sigma \left(\frac{1}{2\omega+3}\right) \left(-3+2\gamma_f\right)^2 \right\}
 +\frac{\varepsilon}{\sigma}+\frac{\gamma_f \, \rho_+}{\left(1-v_+^2\right)\sigma}.
\end{multline}
Finally, Eq.~\eqref{segunda} can be written as follows
\begin{equation}\label{S1}
\dfrac{d \sigma}{d t_+} = -2 \left( \frac{\gamma_f}{R}\right) \dfrac{d
R}{d t_+}  + \frac{\gamma_f \, \rho_{+} v_{+}}{\sqrt{1-v_{+}^2}}\;.
\end{equation}

The evolution of the bubble wall is completely determined, in the
outside coordinates, by the equations (\ref{R1})-(\ref{S1}).
We solved these equations numerically by consider different kind of
matter content for the background, satisfying the stability conditions discussed in Sec.~\ref{sec:estadoestatico}.
From these numerical solutions we found that once the bubble has materialized
in the background of an ES universe in the context of a JBD theory, it grows filling the background space.

\begin{figure}[t]
    \centering
    \subfigure[ ]{\includegraphics[scale=0.5]{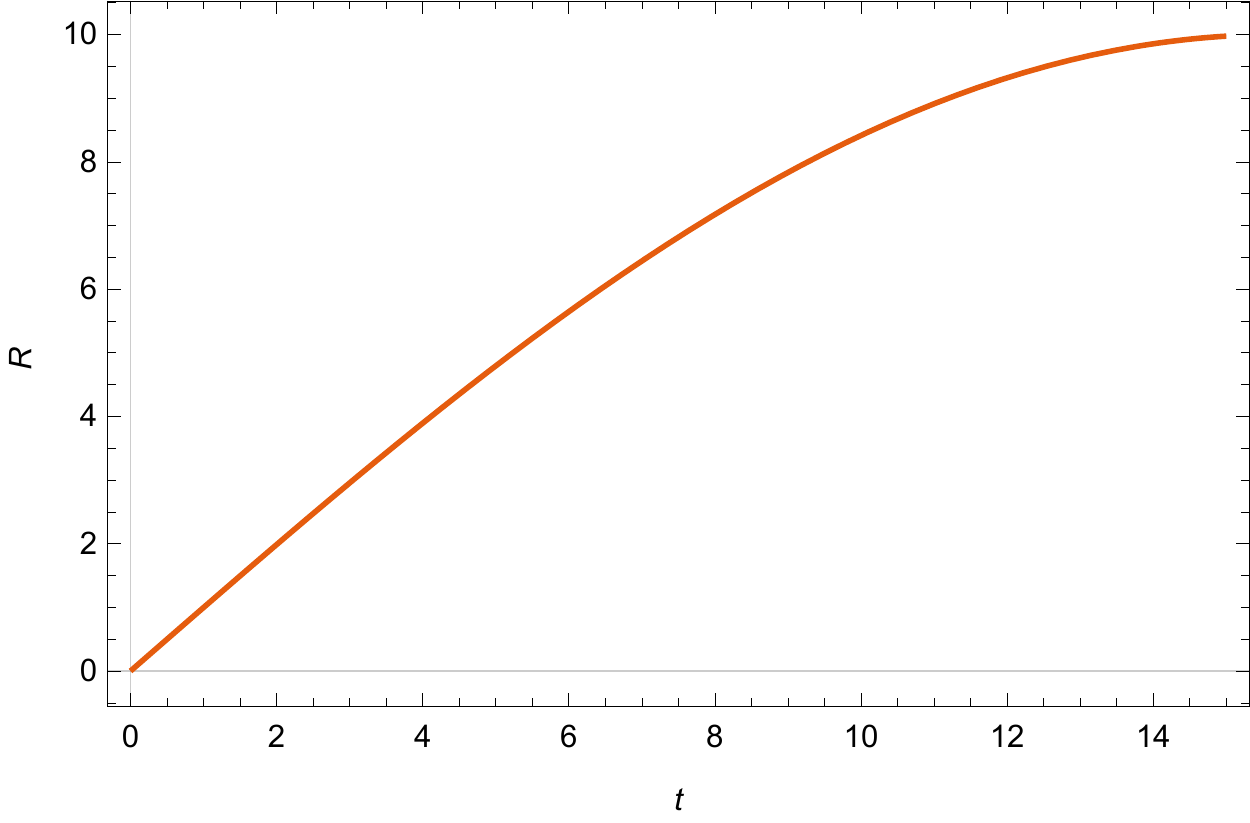}}
    \subfigure[]{\includegraphics[scale=0.53]{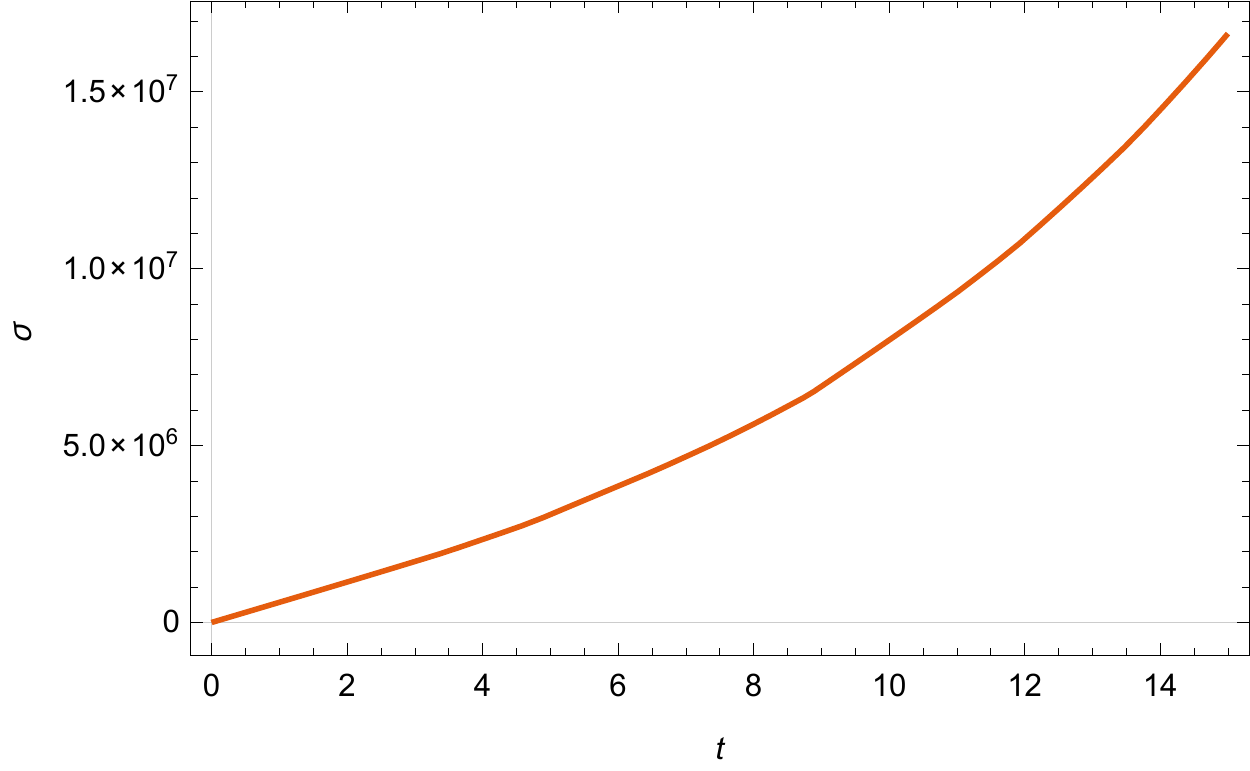}}
\caption{Evolution of the true vacuum bubble for $ \gamma_f = 0.7 $.
(a) Radius of the bubble as a function of time $t_+$. (b) Surface
energy density as a function of time $t_+$.} \label{fig:evolucion1}
\end{figure}

\begin{figure}[t]
    \centering
    \subfigure[ ]{\includegraphics[scale=0.5]{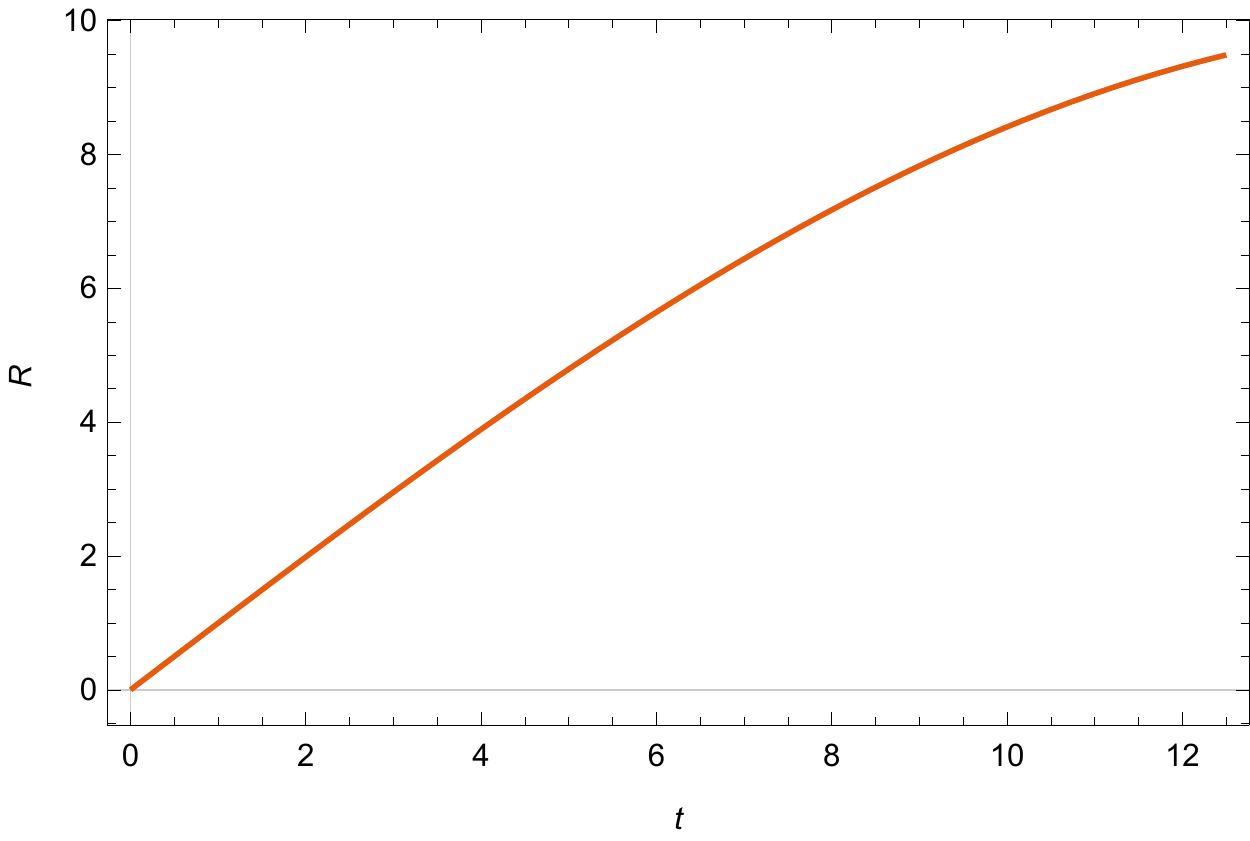}}
    \subfigure[]{\includegraphics[scale=0.53]{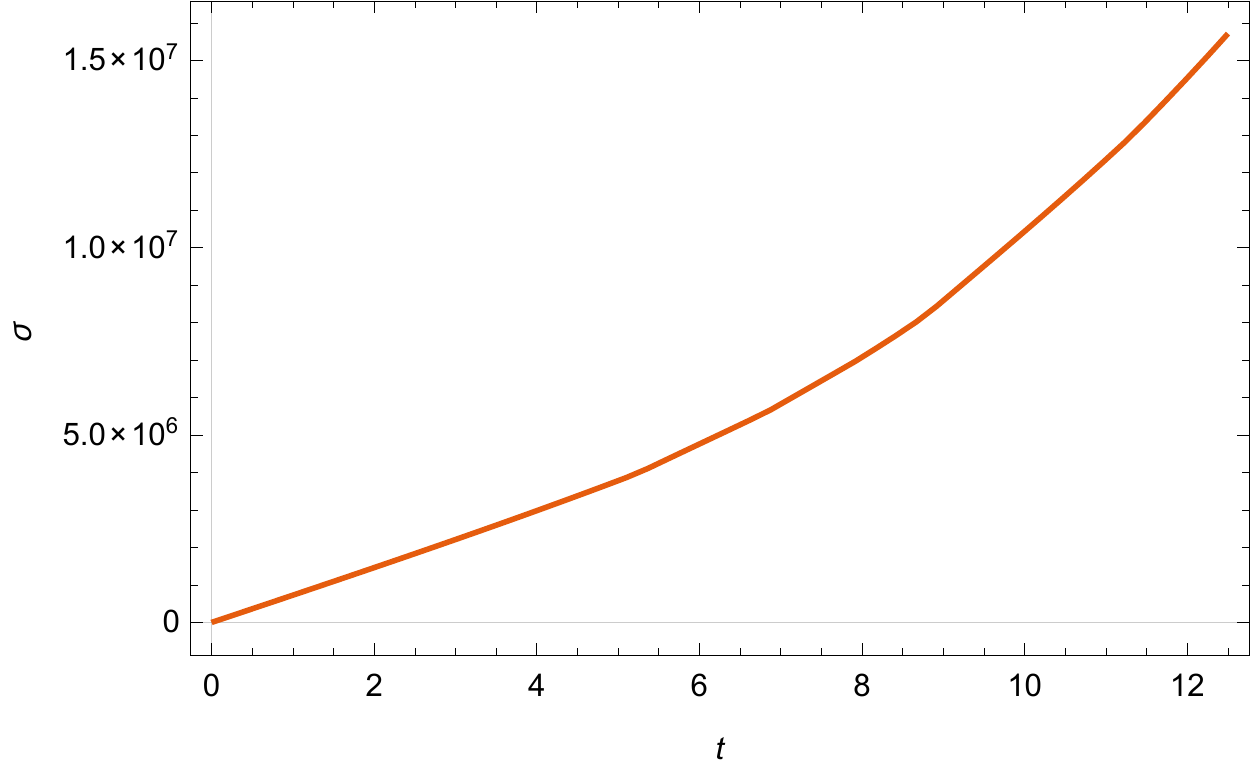}}
\caption{Evolution of the true vacuum bubble for $\gamma_f = 1.3$.
(a) Radius of the bubble as a function of time $t_+$. (b) Surface
energy density as a function of time $t_+$.} \label{fig:evolucion2}
\end{figure}

\begin{figure}[t]
    \centering
    \subfigure[]{\includegraphics[scale=0.5]{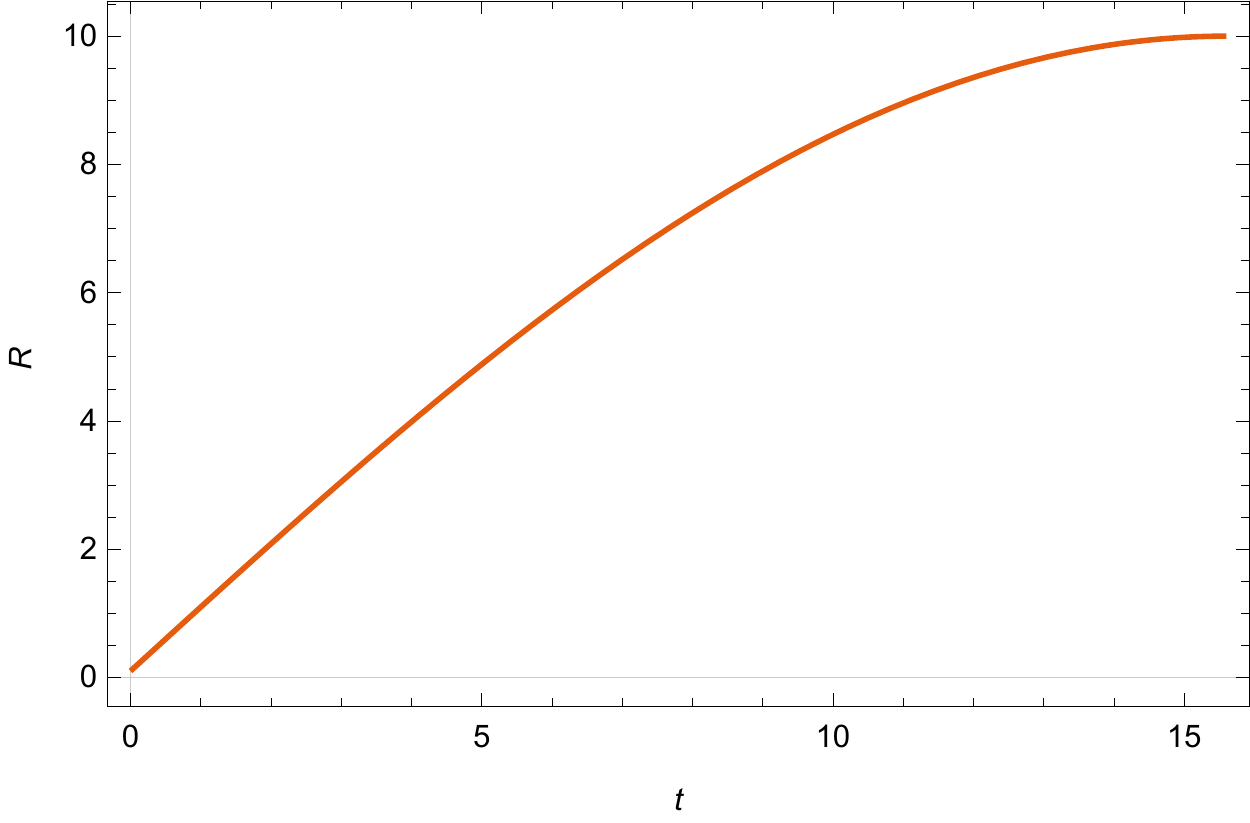}}
    \subfigure[]{\includegraphics[scale=0.53]{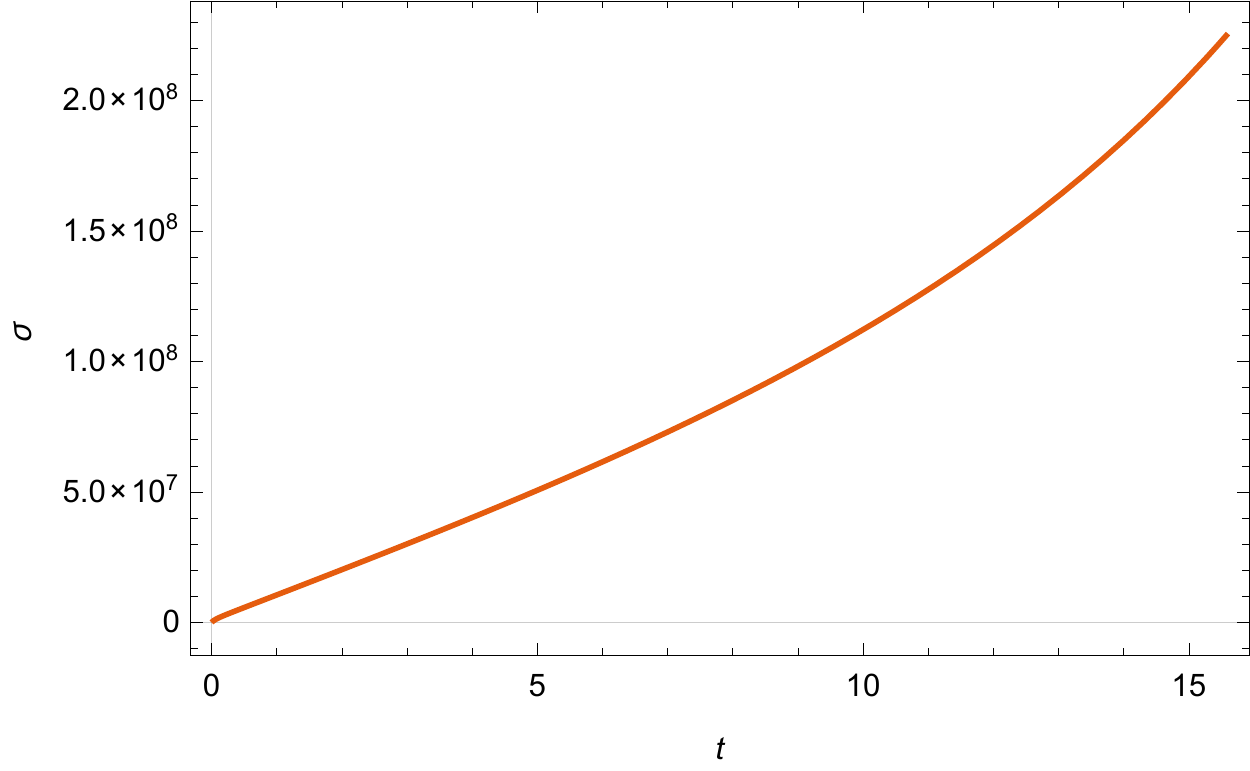}}
\caption{Evolution of the true vacuum bubble for $\gamma_f=1.0$. (a)
Radius of the bubble as a function of time $t_+$. (b) Surface energy
density as a function of time $t_+$.} \label{fig:evolucion3}
\end{figure}

In particular in Figs.~(\ref{fig:evolucion1})-(\ref{fig:evolucion3}) we show three examples.
In the first example we consider $\gamma_f = 0.7 $ and set the
following values for the radius and JBD field of the background
\begin{equation}
    a_0 = 10,\quad \phi_0 = 0.9.
\end{equation}
The results for this case are shown in the Figure (\ref{fig:evolucion1}).

In the second example, we consider $\gamma_f = 1.3$ and the following values
for the radius and the JBD field
\begin{equation}
    a_0=10, \quad \phi_0 =1.3.
\end{equation}
The results for this case are shown in Figure (\ref{fig:evolucion2}).

In the third example, the matter content of the
background is dust ($\gamma_f = 1$) and we consider $a_0=10$ and
$\phi_0 =1.3$. The results are shown in Figure (\ref{fig:evolucion3}).

In all these examples we have assumed the following initial conditions
\begin{equation}
\sigma_{init}=10^{-6}, \quad R_{init}=10^{-3}, \quad
v_{init}=10^{-2},
\end{equation}
and units where $8\pi G = 1$.


From these examples we can note that the bubble of the new face,
once materialized, grows to fill the background space without collapsing.

\section{Conclusions}\label{sec:conclu}

In this paper we study an alternative scheme for an Emergent
Universe scenario called Emergent Universe by tunneling. In this scheme the universe is initially  in a truly
static state supported by a scalar field which is
located in a false vacuum. The universe begins to evolve when, by
quantum tunneling, the scalar field decays into a state of true
vacuum.

The EU by tunneling scheme was originally developed in Ref.~\cite{Labrana:2013kqa}, in the context of General Relativity, where it was concluded that this mechanism is feasible as an EU scheme. Nevertheless, this first model present the problem that the ES solution is classically instable.
The instability of the ES solution ensures that any perturbation,
no matter how small, rapidly force the universe away from the static
state, thereby aborting the EU scenario.

The present work is the natural extension of the idea presented in Ref.~\cite{Labrana:2013kqa}, but where the problem of the classical instability of the static solution is solved by going away from General Relativity and consider a JBD theory.

In particular, in this work we focus our study on the process of tunneling of a
scalar field and the consequent creation and evolution of a bubble of true vacuum in the background of a classically stable Einstein Static universe.
Our principal motivation is the study of new ways of leaving the static period and begin the inflationary regime in the context of Emergent Universe models.

In the first part of the paper, Sect.~\ref{sec:estadoestatico}, we
study an Einstein static universe supported by a scalar field
located in a false vacuum and its stability in the context of a JBD
theory.
Contrary to General Relativity, we found that this static solution
could be stable against isotropic perturbations if some general
conditions are satisfied, see
Eqs.~(\ref{condgamma})-(\ref{vdosprima}).
This modification of the stability behavior has important
consequences for the emergent universe by tunneling scenario, since
it ameliorates the fine-tuning that arises from the fact that the ES
model is an unstable saddle in GR and it improves the preliminary
model studied in Ref.~\cite{Labrana:2011np}.
In this study, for simplicity, we have not considered inhomogeneous
or anisotropic perturbations.
At this respect, the stability of the ES solution under anisotropic, tensor and inhomogeneous scalar perturbations have been studied in the context of JBD theories in Refs.~\cite{delCampo:2009kp, Huang:2014fia}. It was found for theses JBD models that different from General Relativity \cite{Barrow:2003ni} and others modified theories of gravity as $f(R)$ \cite{Seahra:2009ft} or modified Gauss-Bonnet gravity \cite{Huang:2015kca}, that a static universe which is stable against homogeneous perturbations, could be also stable against anisotropic and inhomogeneous perturbations. We expect a similar behavior for our JBD model, were the static universe is supported by a scalar field located in a false vacuum. Then, we expect that in our case the inhomogeneous and anisotropic perturbations do not lead to additional instabilities. Nevertheless, we intend to return to these points in the near future by working an approach similar to that followed in Refs.~\cite{Barrow:2003ni,delCampo:2009kp,inhomogeneous, Huang:2014fia}.

In Sect. \ref{sec:tunel} we study the tunneling process of the
scalar field from the false vacuum to the true vacuum and the
consequent creation of a bubble of true vacuum in the background of
Einstein static universe for a JBD theory. In particular we
determinate the nucleation rate of the true vacuum bubble using the
approaches developed in Ref.~\cite{Holman:1989gh} and previous
results obtained in Ref.~\cite{Labrana:2011np}.

The classical evolution of the bubble after its nucleation is studied in Sect.~\ref{sec:evolucionbur} where we found that once the bubble has materialized in the background of an ES universe, it grows filling the background space.
This demonstrates the viability of our EU model, since
there is the possibility of having an open inflationary universe
inside the bubble.
During this study we consider the gravitational back-reaction of the bubble by using the formalism developed in Ref.~\cite{Sakai:1992ud} applied to a JBD theory. At this respect we found a system of coupled differential equations, which we solved numerically. Three specific examples of these solutions were shown in Sect.~\ref{sec:evolucionbur} concerning to different background material contents.

It is worth to note that once the bubble
has materialized, from conditions \eqref{tresocho}, it follows
that if one of the regions of spacetime separated by the wall is homogeneous,
then the other region is, in general, inhomogeneous \cite{Sakai:1992ud}. Given that in our case the exterior of the bubble is a homogeneous universe, then the interior of the bubble will be, in general,
inhomogeneous. However, since the degree of
inhomogeneity depends on the difference in the energy density of the
interior and the exterior of the bubble, it
is possible in our case to decrease this inhomogeneity by
adjusting the parameters of the static solution as was discussed in Ref.~\cite{Sakai:1992ud}.
Then in our model,
it is possible to study the feasibility of having an open inflationary universe inside the bubble.
Nevertheless, given the similarities, we expect that the behavior inside the bubble of the EU by tunneling, will be similar to the models of single-field open and extended open inflation, as the ones studied in Refs.~\cite{linde, re8, delC1, delC2,Balart:2007je}.
We expect to return to this point in the near future.

\section{Acknowledgements}
 P. L. is supported by Direcci\'on de
Investigaci\'on de la Universidad del B\'{\i}o-B\'{\i}o through
Grants N$^0$ 182707 4/R, and GI 172309/C. H. C. was supported by Direcci\'on de Postgrado de la Universidad del
B\'{\i}o-B\'{\i}o and by Research Assistant Grant of Escuela de Graduados Universidad del B\'{\i}o-B\'{\i}o.


\begin{thebibliography}{99}
%%%%%%%Weinberg,Peebles,Kolb%%%%%%%%%%%%%%
\bibitem{weinberg} S. Weinberg, Gravitation and Cosmology: Principle and Application of the General Relativity, Wiley, NY, 1972; Ch. W. Misner, K. S. Turner, J. A. Wheeler, Gravitation, W. H.: Freeman and Company, SF 1973.
\bibitem{peebles} P. J. E. Peebles, Principles of Physical Cosmology, Princeton University Press 1993; J. A. Peacock, Cosmological Physics, Cambridge University Press, 1998); S. Weinberg, Cosmology, Oxford University Press, USA, 2008.
\bibitem{kolb} E. Kolb and M. Turner, The Early Universe, Addison- Wesley Publishing(1989).

\bibitem{Guth1}  Guth A.,  The inflationary universe:
A possible solution to the horizon and flatness problems, 1981 Phys.
Rev. D {\bf 23} 347.

\bibitem{Albrecht}  Albrecht A. and  Steinhardt P. J.,
Cosmology for grand unified theories with radiatively induced symmetry
breaking, 1982 Phys. Rev. Lett. {\bf 48} 1220.

\bibitem{Linde1}  Linde A. D.,  A new inflationary universe scenario: A possible solution of the horizon,
flatness, homogeneity, isotropy and primordial monopole problems,
1982 Phys. Lett. {\bf 108B} 389.

\bibitem{Linde2}  Linde A. D.,  Chaotic inflation, 1983 Phys. Lett. {\bf 129B} 177.


%%%%%%%%%%%%%%%% singularidad inicial %%%%%%%%%%%%%%%

%\cite{Borde:1993xh}
\bibitem{Borde:1993xh}
  Borde A. and Vilenkin  A.,
  Eternal inflation and the initial singularity,
 1994 Phys.\ Rev.\ Lett.\  {\bf 72} 3305.
  %[arXiv:gr-qc/9312022].
  %%CITATION = PRLTA,72,3305;%%

%\cite{Borde:1997pp}
\bibitem{Borde:1997pp}
  Borde A. and Vilenkin A.,
  Violation of the weak energy condition in inflating spacetimes, 1997
  Phys.\ Rev.\  D {\bf 56} 717.
 % [arXiv:gr-qc/9702019].
  %%CITATION = PHRVA,D56,717;%%

%\cite{Guth:1999rh}
\bibitem{Guth:1999rh}
  A.~H.~Guth,
  ``Eternal inflation,''
  Annals N.\ Y.\ Acad.\ Sci.\  {\bf 950}, 66 (2001)
  %doi:10.1111/j.1749-6632.2001.tb02128.x
  [astro-ph/0101507].


%\cite{Borde:2001nh}
\bibitem{Borde:2001nh}
  ~Borde A., ~Guth A.~H. and ~Vilenkin A.,
  Inflationary space-times are incompletein past directions, 2003
  Phys.\ Rev.\ Lett.\  {\bf 90} 151301.
  %[arXiv:gr-qc/0110012].
  %%CITATION = PRLTA,90,151301;%%


%\cite{Vilenkin:2002ev}
\bibitem{Vilenkin:2002ev}
  ~Vilenkin A.,
  Quantum cosmology and eternal inflation,
  arXiv:gr-qc/0204061.
  %%CITATION = GR-QC/0204061;%%


%%%% EU %%%%%%%%%%%%%%%%%%%

%\cite{Ellis:2002we}
\bibitem{Ellis:2002we}
~Ellis G.~F.~R. and ~Maartens R.,
The emergent universe: Inflationary cosmology with no singularity, 2004
Class.\ Quant.\ Grav.\  {\bf 21} 223.
% [arXiv:gr-qc/0211082].
%%CITATION = CQGRD,21,223;%%

%\cite{Ellis:2003qz}
\bibitem{Ellis:2003qz}
~Ellis G.~F.~R., ~Murugan J. and~Tsagas C.~G.,
The emergent universe: An explicit construction, 2004
Class.\ Quant.\ Grav.\  {\bf 21} 233.
%[arXiv:gr-qc/0307112].
%%CITATION = CQGRD,21,233;%%

%\cite{Mulryne:2005ef}
\bibitem{Mulryne:2005ef}
~Mulryne D.~J., ~Tavakol R., ~Lidsey J.~E. and ~Ellis G.~F.~R.,
An emergent universe from a loop, 2005
Phys.\ Rev.\  D {\bf 71} 123512.
% [arXiv:astro-ph/0502589].
%%CITATION = PHRVA,D71,123512;%%

%%%%%% Fin pot emergente %%%%%%%%%

%\cite{Mukherjee:2005zt}
\bibitem{Mukherjee:2005zt}
~Mukherjee S., Paul B.~C., Maharaj S.~D.  and ~Beesham A.,
Emergent universe in Starobinsky model,
arXiv:gr-qc/0505103.
%

%\cite{Mukherjee:2006ds}
\bibitem{Mukherjee:2006ds}
~Mukherjee S., Paul B.~C., Dadhich N.~K., Maharaj S.~D. and Beesham
A.~,
Emergent universe with exotic matter, 2006
Class.\ Quant.\ Grav.\  {\bf 23} 6927.



%\cite{Banerjee:2007qi}
\bibitem{Banerjee:2007qi}
A.~Banerjee, T. Bandyopadhyay and ~S. Chakraborty, Emergent universe
in brane world scenario, Grav. Cosmol. {\bf 13}, 290 (2007).



%\cite{Nunes:2005ra}
\bibitem{Nunes:2005ra}
~Nunes N.~J.,
Inflation: A graceful entrance from loop quantum cosmology, 2005
Phys.\ Rev.\  D {\bf 72} 103510.

%\cite{Lidsey:2006md}
\bibitem{Lidsey:2006md}
~Lidsey J.~E. and~Mulryne D.~J.,
A graceful entrance to braneworld inflation, 2006
Phys.\ Rev.\  D {\bf 73} 083508.

%\cite{delCampo:2007mp}
\bibitem{delCampo:2007mp}
S.~del Campo, R.~Herrera and P.~Labrana,
``Emergent universe in a Jordan-Brans-Dicke theory,''
JCAP {\bf 0711}  030 (2007).
[arXiv:0711.1559 [gr-qc]].
%%CITATION = JCAPA,0711,030;%%


\bibitem{delCampo:2009kp}
S.~del Campo, R.~Herrera and P.~Labrana,
``On the Stability of Jordan-Brans-Dicke Static Universe,''
JCAP {\bf 0907} (2009) 006.
%doi:10.1088/1475-7516/2009/07/006
[arXiv:0905.0614 [gr-qc]].
%%CITATION = doi:10.1088/1475-7516/2009/07/006;%%

%\cite{delCampo:2010kf}
\bibitem{delCampo:2010kf}
S.~del Campo, E.~Guendelman, R.~Herrera, P.~Labrana,
``Emerging Universe from Scale Invariance,''
JCAP {\bf 1006 } (2010)  026.
[arXiv:1006.5734 [astro-ph.CO]].

%\cite{delCampo:2011mq}
\bibitem{delCampo:2011mq}
S.~del Campo, E.~I.~Guendelman, A.~B.~Kaganovich, R.~Herrera, P.~Labrana,
``Emergent Universe from Scale Invariant Two Measures Theory,''
Phys.\ Lett.\  {\bf B699 } (2011)  211-216.
[arXiv:1105.0651 [astro-ph.CO]].

%%%% Papers Eduardo %%%%%%%%%%

%\cite{Guendelman:2011zza}
\bibitem{Guendelman:2011zza}
  E.~I.~Guendelman,
  %``Non Singular Origin of the Universe and its Present Vacuum Energy Density,''
  Int.\ J.\ Mod.\ Phys.\ A {\bf 26}, 2951 (2011)
%  doi:10.1142/S0217751X11053614
  [arXiv:1103.1427 [gr-qc]].


%\cite{Guendelman:2011fr}
\bibitem{Guendelman:2011fr}
  E.~I.~Guendelman,
  %``Non Singular Origin of the Universe and the Cosmological Constant Problem (CCP),''
  Int.\ J.\ Mod.\ Phys.\ D {\bf 20}, 2767 (2011)
%  doi:10.1142/S0218271811020718
  [arXiv:1105.3312 [gr-qc]].


%\cite{Guendelman:2013dka}
\bibitem{Guendelman:2013dka}
  E.~I.~Guendelman and P.~Labrana,
  %``Connecting The Non-Singular Origin of the Universe, The Vacuum Structure and The Cosmological Constant Problem,''
  Int.\ J.\ Mod.\ Phys.\ D {\bf 22} (2013) 1330018
   [arXiv:1303.7267 [astro-ph.CO]].

%\cite{Guendelman:2014bva}
\bibitem{Guendelman:2014bva}
  E.~Guendelman, R.~Herrera, P.~Labrana, E.~Nissimov and S.~Pacheva,
  %``Emergent Cosmology, Inflation and Dark Energy,''
  Gen.\ Rel.\ Grav.\  {\bf 47} (2015) no.2,  10
    [arXiv:1408.5344 [gr-qc]].

%\cite{Guendelman:2015uca}
\bibitem{Guendelman:2015uca}
  E.~Guendelman, R.~Herrera, P.~Labrana, E.~Nissimov and S.~Pacheva,
  %``Stable emergent Universe – a creation without Big-Bang,''
  Astron.\ Nachr.\  {\bf 336} (2015) no.8/9,  810
  [arXiv:1507.08878 [hep-th]].

%\cite{delCampo:2015yfa}
\bibitem{delCampo:2015yfa}
  S.~del Campo, E.~I.~Guendelman, R.~Herrera and P.~Labrana,
  %``Classically and Quantum stable Emergent Universe from Conservation Laws,''
  JCAP {\bf 1608} (2016) 049.
   [arXiv:1508.03330 [gr-qc]].


%%%%%%%%%%%%% Universos emergentes asintoticos %%%%%


%\cite{Banerjee:2007sg}
\bibitem{Banerjee:2007sg}
A.~Banerjee, T.~Bandyopadhyay, S.~Chakraborty,
%``Emergent Universe in Brane World Scenario with Schwarzschild-de Sitter Bulk,''
Gen.\ Rel.\ Grav.\  {\bf 40}, 1603-1607 (2008).
[arXiv:0711.4188 [gr-qc]].


%\cite{Debnath:2008nu}
\bibitem{Debnath:2008nu}
U.~Debnath,
%``Emergent Universe and Phantom Tachyon Model,''
Class.\ Quant.\ Grav.\  {\bf 25}, 205019 (2008).
[arXiv:0808.2379 [gr-qc]].

%\cite{Paul:2008id}
\bibitem{Paul:2008id}
B.~C.~Paul, S.~Ghose,
%``Emergent Universe Scenario in the Einstein-Gauss-Bonnet Gravity with Dilaton,''
Gen.\ Rel.\ Grav.\  {\bf 42}, 795-812 (2010).
[arXiv:0809.4131 [hep-th]].

%\cite{Beesham:2009zw}
\bibitem{Beesham:2009zw}
A.~Beesham, S.~V.~Chervon, S.~D.~Maharaj,
%``An Emergent universe supported by a nonlinear sigma model,''
Class.\ Quant.\ Grav.\  {\bf 26}, 075017 (2009).
[arXiv:0904.0773 [gr-qc]].


%\cite{Debnath:2011qi}
\bibitem{Debnath:2011qi}
U.~Debnath, S.~Chakraborty,
%``Emergent Universe with Exotic Matter in Brane World Scenario,''
Int.\ J.\ Theor.\ Phys.\  {\bf 50}, 2892-2898 (2011).
[arXiv:1104.1673 [gr-qc]].


%\cite{Mukerji:2011wq}
\bibitem{Mukerji:2011wq}
S.~Mukerji, N.~Mazumder, R.~Biswas, S.~Chakraborty,
%``Emergent scenario and different anisotropic models,''
Int.\ J.\ Theor.\ Phys.\  {\bf 50}, 2708-2719 (2011).
[arXiv:1106.1743 [gr-qc]].

%\cite{Labrana:2013oca}
\bibitem{Labrana:2013oca}
  P.~Labrana,
  %``Emergent universe scenario and the low CMB multipoles,''
  Phys.\ Rev.\ D {\bf 91} (2015) no.8,  083534.
%  doi:10.1103/PhysRevD.91.083534
  [arXiv:1312.6877 [astro-ph.CO]].


%\cite{Huang:2015zma}
\bibitem{Huang:2015zma}
  Q.~Huang, P.~Wu and H.~Yu,
  %``Emergent scenario in the Einstein-Cartan theory,''
  Phys.\ Rev.\ D {\bf 91}, no. 10, 103502 (2015).
%  doi:10.1103/PhysRevD.91.103502
%  [arXiv:1504.05284 [gr-qc]].


%%%%%%%%%%%%%%%%%%%%%%%%%%%%%%%%%%%%%%%%%%%%%%%%%%%

\bibitem{Labrana:2011np}
P.~Labrana,
``Emergent Universe by Tunneling,''
Phys.\ Rev.\ D {\bf 86}, 083524 (2012).
%doi:10.1103/PhysRevD.86.083524
[arXiv:1111.5360 [gr-qc]].
%%CITATION = doi:10.1103/PhysRevD.86.083524;%%
%13 citations counted in INSPIRE as of 07 Sep 2016
\bibitem{Labrana:2013kqa}
P.~Labrana,
``Tunneling and the Emergent Universe Scheme,''
Astrophys.\ Space Sci.\ Proc.\  {\bf 38} (2014) 95.
%doi:10.1007/978-3-319-02063-1-8
%%CITATION = doi:10.1007/978-3-319-02063-1_8;%%
%1 citations counted in INSPIRE as of 04 Jun 2018
\bibitem{Labrana:2014yta}
P.~Labrana,
``The Emergent Universe scheme and Tunneling,''
AIP Conf.\ Proc.\  {\bf 1606} (2014) 38.
%doi:10.1063/1.4891114
[arXiv:1406.0922 [astro-ph.CO]].
%%CITATION = doi:10.1063/1.4891114;%%
%1 citations counted in INSPIRE as of 04 Jun 2018

\bibitem{Eddington}
A.~S.~Eddington, Mon.\ Not.\ Roy.\ Astron.\ Soc.\  {\bf 90}, 668
(1930).
%%CITATION = MNRAA,90,668;%%

%\cite{Harrison:1967zz}
\bibitem{Harrison:1967zz}
E.~R.~Harrison,
%``Normal Modes Of Vibrations Of The Universe,''
Rev.\ Mod.\ Phys.\  {\bf 39}, 862 (1967).
%%CITATION = RMPHA,39,862;%%
%%%%%%% Consideraciones de entropia

%\cite{Gibbons:1987jt}
\bibitem{Gibbons:1987jt}
~Gibbons G.~W.,
The entropy and stability of the universe, 1987
Nucl.\ Phys.\  B {\bf 292} 784 .
%%CITATION = NUPHA,B292,784;%%

%\cite{Gibbons:1988bm}
\bibitem{Gibbons:1988bm}
~Gibbons G.~W.,
Sobolev's inequality, Jensen's theorem and the mass and entropy of the
universe, 1988
Nucl.\ Phys.\  B {\bf 310} 636.
%%CITATION = NUPHA,B310,636;%%

%\cite{Barrow:2003ni}
\bibitem{Barrow:2003ni}
J.~D.~Barrow, G.~F.~R.~Ellis, R.~Maartens and C.~G.~Tsagas,
``On the stability of the Einstein static universe,''
Class.\ Quant.\ Grav.\  {\bf 20}, L155 (2003).
%doi:10.1088/0264-9381/20/11/102
[gr-qc/0302094].
%%CITATION = doi:10.1088/0264-9381/20/11/102;%%
%94 citations counted in INSPIRE as of 09 Mar 2017


%\cite{Huang:2014fia}
\bibitem{Huang:2014fia}
  H.~Huang, P.~Wu and H.~Yu,
  %``Stability of the Einstein static universe in the Jordan-Brans-Dicke theory,''
  Phys.\ Rev.\ D {\bf 89}, no. 10, 103521 (2014).
%  doi:10.1103/PhysRevD.89.103521


\bibitem{Jbd}  Jordan P.,  The present state of Dirac's cosmological
hypothesis, 1959 Z.Phys. {\bf 157} 112;
Brans C.~ and ~Dicke R.~H.,
Mach's principle and a relativistic theory of gravitation, 1961
Phys.\ Rev.\  {\bf 124} 925.
%%CITATION = PHRVA,124,925;%%

%%%%%%%%%%%%%%%%%%%%%%% Kaluza-klein %%%%%%%%%%%
%\cite{Freund:1982pg}
\bibitem{Freund:1982pg}
~Freund P.~G.~O.,
Kaluza-Klein cosmologies, 1982
Nucl.\ Phys.\  B {\bf 209} 146.
%%CITATION = NUPHA,B209,146;%%

%\cite{Appelquist:1987nr}
\bibitem{Appelquist:1987nr}
Appelquist T., Chodos A. and ~Freund P.~G.~O.,
%
\textit{Modern Kaluza-Klein theories} ( 1987 Addison-Wesley, Redwood
City).
%
%%%%%%%%%%%%%%% Superstring %%%%%%%%%%%%%%%%%%%%%%

%\cite{Fradkin:1984pq}
\bibitem{Fradkin:1984pq}
~Fradkin E.~S. and Tseytlin A.~A.,
Effective field theory from quantized strings, 1985
Phys.\ Lett.\  B {\bf 158} 316.
%%CITATION = PHLTA,B158,316;%%

%\cite{Fradkin:1985ys}
\bibitem{Fradkin:1985ys}
~Fradkin E.~S. and ~Tseytlin A.~A.,
Quantum string theory effective action, 1985
Nucl.\ Phys.\  B {\bf 261} 1.
%%CITATION = NUPHA,B261,1;%%

%\cite{Callan:1985ia}
\bibitem{Callan:1985ia}
Callan C.~G., Martinec E.~J., Perry M.~J.~ and ~Friedan D.,
Strings in background fields, 1985
Nucl.\ Phys.\  B {\bf 262} 593.
%%CITATION = NUPHA,B262,593;%%

%\cite{Callan:1986jb}
\bibitem{Callan:1986jb}
CallanC.~G., Klebanov I.~R.~ and Perry M.~J.~,
String theory effective actions, 1986
Nucl.\ Phys.\  B {\bf 278} 78.
%%CITATION = NUPHA,B278,78;%%

%\cite{Green:1987sp}
\bibitem{Green:1987sp}
~Green M.~B., ~Schwarz J.~H. and ~Witten E., \textit{Superstring theory}
(1987 Cambridge, Uk: Univ. Pr., Cambridge Monographs On
Mathematical Physics).

\bibitem{Coleman:1977py}
S.~R.~Coleman,
``The Fate of the False Vacuum. 1. Semiclassical Theory,''
Phys.\ Rev.\ D {\bf 15}, 2929 (1977)
Erratum: [Phys.\ Rev.\ D {\bf 16}, 1248 (1977)].
%doi:10.1103/PhysRevD.15.2929, 10.1103/PhysRevD.16.1248
%%CITATION = doi:10.1103/PhysRevD.15.2929, 10.1103/PhysRevD.16.1248;%%
%1667 citations counted in INSPIRE as of 27 Mar 2017


\bibitem{Coleman:1980aw}
S.~R.~Coleman and F.~De Luccia,
``Gravitational Effects on and of Vacuum Decay,''
Phys.\ Rev.\ D {\bf 21}, 3305 (1980).
%doi:10.1103/PhysRevD.21.3305
%%CITATION = doi:10.1103/PhysRevD.21.3305;%%
%1164 citations counted in INSPIRE as of 27 Mar 2017

%%%%%%% Universos abiertos %%%%%%%

\bibitem{linde} A. Linde, Phys. Rev. D {\bf 59}, 023503 (1998).

\bibitem{re8}A. Linde, M. Sasaki and T. Tanaka, Phys. Rev. D {\bf
    59}, 123522 (1999).

\bibitem{delC1} S. del Campo and R. Herrera, Phys. Rev. D {\bf 67}, 063507 (2003).

\bibitem{delC2} S. del Campo, R. Herrera and J. Saavedra, Phys. Rev. D {\bf 70}, 023507 (2004).

%\cite{Balart:2007je}
\bibitem{Balart:2007je}
L.~Balart, S.~del Campo, R.~Herrera, P.~Labrana, J.~Saavedra,
``Tachyonic open inflationary universes,''
Phys.\ Lett.\  {\bf B647}, 313-319 (2007).
%[gr-qc/0703026 [GR-QC]].
%
%
%%%%%%%% Condiciones iniciales

\bibitem{Antoniadis}
I. Antoniadis, C. Bachas, J. Ellis and D.V. Nanopolous, Phys. Lett.
B 211, 4 (1988).

\bibitem{Tryon}
E.P. Tryon, Nature(London) 246, 396 (1973).

\bibitem{Vilenkin-cre}
A. Vilenkin, Phys. Rev. D 32, 10 (1985).

%\cite{Mithani}
\bibitem{Mithani}
  A.~T.~Mithani and A.~Vilenkin,
  %``Collapse of simple harmonic universe,''
  JCAP {\bf 1201}, 028 (2012);
%  [arXiv:1110.4096 [hep-th]];
%
%\cite{Mithani:2014jva}
%\bibitem{Mithani:2014jva}
  A.~T.~Mithani and A.~Vilenkin,
  %``Instability of an emergent universe,''
  JCAP {\bf 1405}, 006 (2014);
 % [arXiv:1403.0818 [hep-th]].
%
%\cite{Mithani:2014toa}
%\bibitem{Mithani:2014toa}
  A.~T.~Mithani and A.~Vilenkin,
  %``Stabilizing oscillating universes against quantum decay,''
JCAP {\bf 1507}, no. 07, 010 (2015).
%  [arXiv:1407.5361 [hep-th]].
%%%%%%%%%%%%%%%%%%%%%%%%%%%%%%%%%%%%%%
%
\bibitem{Green:2012oqa}
M.~B.~Green, J.~H.~Schwarz and E.~Witten,
``Superstring Theory Vol. 1 : 25th Anniversary Edition,''
doi:10.1017/CBO9781139248563
%%CITATION = doi:10.1017/CBO9781139248563;%%
%4 citations counted in INSPIRE as of 29 May 2018

%\cite{Simon:2009nb}
\bibitem{Simon:2009nb}
D.~Simon, J.~Adamek, A.~Rakic and J.~C.~Niemeyer,
%``Tunneling and propagation of vacuum bubbles on dynamical backgrounds,''
JCAP {\bf 0911} (2009) 008.
%doi:10.1088/1475-7516/2009/11/008
[arXiv:0908.2757 [gr-qc]].
%%CITATION = doi:10.1088/1475-7516/2009/11/008;%%
%15 citations counted in INSPIRE as of 05 Jun 2018

%\cite{Fischler:2007sz}
\bibitem{Fischler:2007sz}
W.~Fischler, S.~Paban, M.~Zanic and C.~Krishnan,
%``Vacuum bubble in an inhomogeneous cosmology: A Toy model,''
JHEP {\bf 0805} (2008) 041.
%doi:10.1088/1126-6708/2008/05/041
[arXiv:0711.3417 [hep-th]].
%%CITATION = doi:10.1088/1126-6708/2008/05/041;%%
%21 citations counted in INSPIRE as of 05 Jun 2018

\bibitem{Holman:1989gh}
R.~Holman, E.~W.~Kolb, S.~L.~Vadas, Y.~Wang and E.~J.~Weinberg,
``False Vacuum Decay in Jordan-brans-dicke Cosmologies,''
Phys.\ Lett.\ B {\bf 237}, 37 (1990).
%doi:10.1016/0370-2693(90)90457-H
%%CITATION = doi:10.1016/0370-2693(90)90457-H;%%
%31 citations counted in INSPIRE as of 07 Sep 2016
%
%%%%% Nuevos %%%%%%
%\cite{Lee:2008hz}
\bibitem{Lee:2008hz}
  B.~H.~Lee and W.~Lee,
  %``Vacuum bubbles in a de Sitter background and black hole pair creation,''
  Class.\ Quant.\ Grav.\  {\bf 26}, 225002 (2009)
  doi:10.1088/0264-9381/26/22/225002
  [arXiv:0809.4907 [hep-th]].
  %%CITATION = doi:10.1088/0264-9381/26/22/225002;%%
  %29 citations counted in INSPIRE as of 06 Dec 2018

%\cite{Kim:2010yr}
\bibitem{Kim:2010yr}
  H.~Kim, B.~H.~Lee, W.~Lee, Y.~J.~Lee and D.~h.~Yeom,
  %``Nucleation of vacuum bubbles in Brans-Dicke type theory,''
  Phys.\ Rev.\ D {\bf 84}, 023519 (2011)
  doi:10.1103/PhysRevD.84.023519
  [arXiv:1011.5981 [hep-th]].
  %%CITATION = doi:10.1103/PhysRevD.84.023519;%%
  %14 citations counted in INSPIRE as of 06 Dec 2018


\bibitem{Berezin:1987bc}
V.~A.~Berezin, V.~A.~Kuzmin and I.~I.~Tkachev,
%``Dynamics of Bubbles in General Relativity,''
Phys.\ Rev.\ D {\bf 36}, 2919 (1987).
%doi:10.1103/PhysRevD.36.2919
%%CITATION = doi:10.1103/PhysRevD.36.2919;%%
%227 citations counted in INSPIRE as of 24 Sep 2016


%%% Burbujas %%%%%%%%%%%%%%%%%%

\bibitem{Sakai:1992ud}
N.~Sakai and K.~i.~Maeda,
``Bubble dynamics in generalized Einstein theories,''
Prog.\ Theor.\ Phys.\  {\bf 90}, 1001 (1993).
%doi:10.1143/PTP.90.1001
%%CITATION = doi:10.1143/PTP.90.1001;%%
%15 citations counted in INSPIRE as of 07 Sep 2016

\bibitem{Israel:1966rt}
W.~Israel,
``Singular hypersurfaces and thin shells in general relativity,''
Nuovo Cim.\ B {\bf 44S10}, 1 (1966)
[Nuovo Cim.\ B {\bf 44}, 1 (1966)]
Erratum: [Nuovo Cim.\ B {\bf 48}, 463 (1967)].
\bibitem{darmois} Darmois, G.``Memorial des Sciences Mathematiques'' Fasc. 25, Gauthier-Villars (1927)

\bibitem{suffern}
K. G. Suffern, J. of Phys. A15 (1982), 1599.

%%%%%%%%%%%%%%%%%%%%%%%%%%%%%%%%%%%

%\cite{Bruni:1992dg}
\bibitem{inhomogeneous}
  M.~Bruni, P.~K.~S.~Dunsby and G.~F.~R.~Ellis,
  %``Cosmological perturbations and the physical meaning of gauge invariant variables,''
  Astrophys.\ J.\  {\bf 395}, 34 (1992);
%
%\cite{Dunsby:1991xk}
%\bibitem{Dunsby:1991xk}
  P.~K.~S.~Dunsby, M.~Bruni and G.~F.~R.~Ellis,
  %``Covariant Perturbations in a multifluid cosmological medium,''
  Astrophys.\ J.\  {\bf 395}, 54 (1992);
%  doi:10.1086/171630
%
 %\cite{Bruni:1991kb}
%\bibitem{Bruni:1991kb}
  M.~Bruni, G.~F.~R.~Ellis and P.~K.~S.~Dunsby,
  %``Gauge invariant perturbations in a scalar field dominated universe,''
  Class.\ Quant.\ Grav.\  {\bf 9}, 921 (1992);
%  doi:10.1088/0264-9381/9/4/010
%
 %\cite{Dunsby:1998hd}
%\bibitem{Dunsby:1998hd}
  P.~K.~S.~Dunsby, B.~A.~C.~C.~Bassett and G.~F.~R.~Ellis,
  %``Covariant analysis of gravitational waves in a cosmological context,''
  Class.\ Quant.\ Grav.\  {\bf 14}, 1215 (1997).

%%%%%%%%%% Nuevos %%


%\cite{Seahra:2009ft}
\bibitem{Seahra:2009ft}
  S.~S.~Seahra and C.~G.~Boehmer,
  %``Einstein static universes are unstable in generic f(R) models,''
  Phys.\ Rev.\ D {\bf 79}, 064009 (2009).
%  doi:10.1103/PhysRevD.79.064009
%  [arXiv:0901.0892 [gr-qc]].


%\cite{Huang:2015kca}
\bibitem{Huang:2015kca}
  H.~Huang, P.~Wu and H.~Yu,
  %``Instability of the Einstein static universe in modified Gauss-Bonnet gravity,''
  Phys.\ Rev.\ D {\bf 91}, no. 2, 023507 (2015).
%  doi:10.1103/PhysRevD.91.023507


\end{thebibliography}
\end{document}